\documentclass{article}
\usepackage{slashed,verbatim,subfigure}
\usepackage[numbers,sort&compress]{natbib}
\usepackage{amsmath}
\usepackage{amssymb}
\usepackage{authblk}
\usepackage{appendix}
\usepackage{graphicx}
\usepackage{dcolumn}
\usepackage{bm}
\usepackage[colorlinks=true,linktocpage=true,
linkcolor=blue,citecolor=blue]{hyperref}
\usepackage[a4paper]{geometry}
\newcommand{\be}{\begin{eqnarray}}
\newcommand{\ee}{\end{eqnarray}}
\newcommand{\ec}{\sigma_{\rm el}}

\begin{document}
\large
\title{\bf{Analyzing the transport coefficients and observables of a rotating QGP medium in kinetic theory framework with a novel approach to the collision integral}}
\author[1]{Shubhalaxmi Rath\thanks{shubhalaxmirath@gmail.com}}
\author[2]{Sadhana Dash\thanks{sadhana@phy.iitb.ac.in}}
\affil[1]{Instituto de Alta Investigaci\'{o}n, Universidad de Tarapac\'{a}, Casilla 7D, Arica, Chile}
\affil[2]{Department of Physics, Indian Institute of Technology Bombay, Mumbai 400076, India}
\date{}
\maketitle

\begin{abstract}
In the present work, we have studied how the rotation of the QGP medium affects the transport coefficients and observables in heavy ion collisions. For the noncentral collisions, although most of the angular momentum gets carried away by the spectators, there still remains a finite angular momentum with a finite range of angular velocity, which thus incites rotation in the produced matter. As a result, various properties of the QGP medium including its transport properties are most likely to be modulated by the rotation. We have calculated the transport coefficients and observables, such as the electrical conductivity, the thermal conductivity, the Knudsen number, the elliptic flow, the specific heat at constant pressure, the specific heat at constant volume, the trace anomaly, the thermal diffusion constant and the isothermal compressibility using the kinetic theory to see the effect of rotation on them. In particular, we have used the novel relaxation time approximation for the collision integral in the relativistic Boltzmann transport equation to derive the transport coefficients and compared them with their values in the relaxation time approximation within the kinetic theory approach in conjunction with the finite angular velocity. We have found that the emergence of angular velocity enhances the flow of charge and heat in the medium. Further, as 
compared to the relaxation time approximation, the electrical and the thermal conductivities have smaller values in the novel relaxation time approximation and these differences between the conductivities in the said approximations are more pronounced at high temperatures than at low temperatures. Furthermore, all the aforesaid observables are found to be sensitive to the rotation of the QGP medium. 

\end{abstract}

\newpage

\section{Introduction}
The possibility of creating the extreme state of matter that existed in the beginning of the universe, {\em i.e.} quark-gluon plasma (QGP) in ultrarelativistic heavy ion collisions at Relativistic Heavy Ion Collider (RHIC) and Large Hadron Collider (LHC) has elevated the exploration and understanding of the properties of 
such matter. In addition, the study of the evolution of the primordial matter, its equation of state and the phase structure within the quantum chromodynamics (QCD) at finite temperature, density and other extreme conditions are some of the important goals of the heavy ion collisions. There exist evidences that, in the noncentral heavy ion collisions, the nuclei colliding at ultrarelativistic energies could possess a large initial angular momentum of the order $J \propto b\sqrt{s_{NN}}$, where $b$ and $\sqrt{s_{NN}}$ are the impact parameter and the center-of-mass energy, respectively. Although a large part of the angular momentum is carried away by the spectators, there still a significant fraction of the angular momentum varying between $10^3 \hbar$ and $10^5 \hbar$ with angular velocity in the range 0.01 GeV - 0.1 GeV or even more is retained in the interaction zone due to the inhomogeneity of the colliding nuclei in the transverse plane, which further gets transferred to the produced fireball \cite{Deng:PRC93'2016,Jiang:PRC94'2016,Wang:PRD99'2019,Wei:PRD105'2022}. Thus, the larger overlap region between the colliding nuclei or impact parameter and higher collision energy may result into larger angular momentum of the QGP. The fireball or the QGP may sustain the large angular momentum for a longer time due to the conservation of the total angular momentum. Thus, QGP may be perceived as a rotating system and accordingly, its properties could be altered. Consequently, the initial angular momentum of the QGP may give rise to conspicuous observable effects in heavy ion collisions, for example, global polarization of the thermal photons, dileptons and final hadrons with spin, enhancement of the elliptic flow and broadening of the transverse momentum spectra \cite{Liang:PLB629'2005,Becattini:PRC77'2008}. Like the rotating QGP medium, other systems where rotating matter exists are the spinning neutron stars \cite{Berti;MNRAS358'2005} and the nonrelativistic bosonic cold atoms \cite{Fetter:RMP81'2009}. Rotating QGP medium with finite magnetic field also induces some interesting phenomenological effects, such as chiral vortical effect \cite{Son:PRL103'2009,Kharzeev:PRL106'2011}, chiral vortical wave \cite{Jiang:PRD92'2015}, chiral magnetic effect \cite{Fukushima:PRD78'2008,Kharzeev:NPA803'2008}, chiral magnetic wave \cite{Kharzeev:PRD83'2011,Burnier:PRL107'2011}, spin alignment of vector mesons \cite{Liang:PLB629'2005}, emission of circularly polarized photons \cite{Ipp:PLB666'2008} and spin polarization of observed particles in heavy ion collisions \cite{Betz:PRC76'2007,Becattini:AP323'2008,Becattini:AP338'2013,Florkowski:PPNP108'2019, Bhadury:PRL129'2022,Victor:PRD106'2022}. 

In a system, angular momentum exhibits the vorticity, and the coupling between the rotational motion and the quantum spin can result into polarization of hadrons, or in other words, the polarization of emitted hadrons in heavy ion collisions can be treated as a signature of vorticity in the produced matter. The initial conditions of the collisions, {\em i.e.} the energy and the impact parameter mainly decide the amount of polarization of observable particles. A review on the progress of the theoretical and experimental activities on the vortical behavior of the QGP is found in ref. \cite{Becattini:ARNPS70'2020}. The relativistic rotational effect could also influence the thermodynamics and the critical temperature of hadronic phase transition to QGP in QCD. The formulation of the finite temperature field theory in a rotating system can be found in ref. \cite{Vilenkin:PRD21'1980}. Most of the observations in different approaches, such as the Nambu-Jona-Lasinio model \cite{Chernodub:JHEP1701'2017,Wang:PRD99'2019}, the local density approximation with the boundary condition \cite{Ebihara:PLB764'2017}, the hadron resonance gas model \cite{Fujimoto:PLB816'2021} and the holographic QCD approach \cite{Chen:JHEP2107'2021} had found that the rotation decreases the critical temperature, whereas an increase of the critical temperature was reported in lattice simulations \cite{Braguta:JETPL112'2020,Braguta:PRD103'2021}. According to the study using the Nambu-Jona-Lasinio model, it was estimated that the rotation also decreases the critical temperature of chiral symmetry breaking 
due to the suppression of chiral condensate in the similar regime \cite{Jiang:PRL117'2016}. Thus, 
like the strong magnetic field, high temperature and large chemical potential, the rapid rotation of QGP matter can be treated as a kind of extreme condition. Out of different extreme conditions, magnetic field directly 
affects the quark degrees of freedom and indirectly affects the gluon degrees of freedom only through the quark loops, unlike the temperature that affects all degrees of freedom in QCD. The rotation could also affect quark as well as gluon degrees of freedom. 

Recently, a lot of effort has been put into the investigation of various features of magnetized and fast rotating QCD matter due to the emergence of extremely strong magnetic fields and large angular momenta 
in the noncentral events of heavy ion collisions. Both magnetic field and rotation would adequately alter the properties of hot and dense matter. Studies on the rotating free fermion states were carried 
out in references \cite{Iyer:PRD26'1982,Becattini:AP323'2008,Becattini:PRD84'2011,Victor:PLB734'2014,Victor:PRD93'2016}, whereas the system of interacting fermions was studied in references \cite{Chen:PRD93'2016,Jiang:PRL117'2016,Ebihara:PLB764'2017} using effective field theoretical models, in references \cite{McInnes:NPB887'2014,McInnes:NPB911'2016} using the holographic model and in ref. \cite{Chernodub:JHEP1701'2017} using the Nambu-Jona-Lasinio model. Further, the impact of vorticity on the dilepton production rate was investigated in ref. \cite{Singh:PRD100'2019}, where it was reported that the dilepton yield gets suppressed due to the presence of vorticity and the degree of suppression depends on the initial value of vorticity. The effect of magnetic field on different properties of the thermal QCD medium has been extensively studied using different methods \cite{Hees:PRC84'2011,Nam:PRD87'2013,Tuchin:PRC88'2013,Mamo:JHEP1308'2013,Hattori:PRD94'2016,
Fukushima:PRD93'2016,Rath:JHEP1712'2017,Feng:PRD96'2017,Hattori:PRD96'2017,Fukushima:PRL120'2018,Li:PRD97'2018,
Denicol:PRD98'2018,Rath:PRD100'2019,Bandyopadhyay:PRD100'2019,Karmakar:PRD99'2019,
Kurian:EPJC79'2019,Rath:EPJA55'2019,Rath:PRD102'2020,Rath:EPJC80'2020,Rath:EPJC81'2021,Rath:EPJC82'2022,
Rath:EPJA59'2023,Shaikh:PRD108'2023}, while the effect of rotation attracts growing interests recently. However, works on the thermodynamic and transport properties of fast rotating QGP medium are still relatively few in the literature. 

In this work, we intend to study the transport properties related to charge and heat, and some associated observables for the QGP medium subjected to the extreme conditions of rapid rotation and 
high temperature. It would be interesting to see how the rapid rotation modulates the flow of charge 
and heat in QGP medium. In addition, the observables, such as the Knudsen number, elliptic flow, specific heat
at constant pressure, specific heat at constant volume, trace anomaly, thermal diffusion constant 
and isothermal compressibility are also most likely to be affected by the rotation of the QGP medium. Consequently, it can be expected that this study on the transport phenomena and observables of QGP medium with rotation will provide deeper insight into the properties of the strongly interacting system. We determine the transport coefficients, such as the electrical conductivity and the thermal conductivity using the relativistic Boltzmann transport equation in the novel relaxation time approximation within the kinetic theory approach. The relaxation time approximation may defy the local particle number conservation in the medium, because the charge is not conserved instantaneously but only on the average basis over a cycle. However, the validity of the conservation laws can be satisfied by using the novel relaxation time approximation. In this method, the collision integral takes a modified form to ensure the validity of the energy-momentum conservation laws \cite{Cercignani:BOOK'2002,Rocha:PRL127'2021}. Thus the novel relaxation time approximation preserves the fundamental properties of the collision associated with the microscopic conservation laws and is imperative when considering the momentum-dependent relaxation time. In addition, the rotational effect is incorporated through the parton distribution functions. Further, the interactions among partons are included through their thermal masses in the quasiparticle model. In this model, the quark-gluon plasma is treated as a medium consisting of the thermally massive quasiparticles. 

The present paper is organized as follows. In section 2, the electrical and the thermal conductivities of a rotating QGP medium are studied by calculating them using the relativistic Boltzmann transport equation 
in the novel relaxation time approximation within the kinetic theory approach. The effects of 
rotation on the observables, such as the Knudsen number, elliptic flow, specific heat at constant pressure, specific heat at constant volume, trace anomaly, thermal diffusion constant and isothermal compressibility 
are also studied. The results on the aforesaid conductivities and observables are discussed in section 3. In 
section 4, the results of this work are concluded. 

\section{Rotating QGP medium}
The emergence of rotation in a system could affect its various properties including the transport properties. In ref. \cite{Becattini:AP325'2010}, the phase-space distribution function for a relativistic rotating gas consisting of massive particles with spin $S$ in the Boltzmann limit had been formulated. It can be generalized to the case of rotating QGP medium consisting of the thermally massive quarks and gluons within the quasiparticle model (outside the quasiparticle model, {\em i.e.} with bare masses, the generalization might not be valid). The generalization of the distribution functions within the quasiparticle model is plausible, because it is well-known that, in the high temperature QCD regime, the thermally massive quark and gluon distribution functions can be approximated to the Boltzmann distributions. Thus, in a rotating QGP medium, the phase-space distribution functions of partons get affected. In such rotating system with angular velocity $\omega$, the density operator can be written as
\be\label{D.O.}
\hat{\rho}=\frac{1}{Z}e^{\beta\left(-\hat{H}+\mu\hat{N}+\omega\hat{J}\right)}P_V
,\ee
where $\hat{H}$, $\hat{N}$ and $\hat{J}$ represent the Hamiltonian operator, the number operator and the angular momentum operator, respectively. Here, $T=\beta^{-1}$ and $P_V$ denotes the projector onto the localized states $|h_V\rangle$, {\em i.e.} $P_V=\sum_{h_V}|h_V\rangle\langle h_V|$, which forms a complete set of quantum states for the system in a finite region $V$. The partition function, $Z$ of the rotating system is expressed as
\be\label{P.F.}
Z=Tr\left[e^{\beta\left(-\hat{H}+\mu\hat{N}+\omega\hat{J}\right)}P_V\right]
.\ee
In the Boltzmann limit, it is possible to write the partition function \eqref{P.F.} in terms of a product 
of the single particle partition functions. Thus, eq. \eqref{P.F.} can be determined by computing the 
matrix elements of operators $\hat{h}$, $\hat{n}$ and $\hat{j}$, suitable for single particles \cite{Becattini:AP325'2010} as
\be\label{M.E.}
\langle p,\tau|e^{\beta\left(-\hat{h}+\mu\hat{n}+\omega\hat{j}\right)}P_V|p,\sigma \rangle=\left(e^{\beta\left(u_\mu p^\mu+a\right)}+b\right)^{-1}\langle p,\tau|e^{\beta\omega\hat{j}}P_V|p,\sigma \rangle
,\ee
where $a=-\mu$, $+\mu$ and $0$ for quarks, antiquarks and gluons, respectively, and $b=+1$ for quarks as well as for antiquarks and $b=-1$ for gluons. In the above equation, $\tau$ and $\sigma$ represent the 
polarization states. The matrix element appearing on the right-hand side of eq. \eqref{M.E.} can be determined by the analytical continuation to the imaginary $\omega$ values. In this process, $\beta\omega$ is replaced by $-i\phi$ through the rotation $R_{\hat{\omega}}(\phi)$ around the axis $\omega$ by an angle $\phi$. Now, the matrix element on the right-hand side of eq. \eqref{M.E.} can be expanded \cite{Becattini:AP325'2010} as
\be\label{M.E.(1)}
\nonumber\langle p,\tau|e^{\beta\omega\hat{j}}P_V|p,\sigma \rangle &=& \langle p,\tau|R_{\hat{\omega}}(\phi) P_V|p,\sigma \rangle \\ &=& \sum_{\sigma^\prime}\int d^3{\rm p^\prime} \langle p,\tau|R_{\hat{\omega}}(\phi)|p^\prime,\sigma^\prime \rangle \langle p^\prime,\sigma^\prime|P_V|p,\sigma \rangle
.\ee
The matrix element of the representation of a rotation involves a Dirac delta function and a Wigner rotation matrix, {\em i.e.}, 
\be\label{M.E.(2)}
\langle p,\tau|R_{\hat{\omega}}(\phi)|p^\prime,\sigma^\prime \rangle=\delta^3\left(p-R_{\hat{\omega}}(\phi)(p^\prime)\right)D^S\left(\left[R_{\hat{\omega}}(\phi)(p^\prime)\right]^{-1}R_{\hat{\omega}}(\phi)\left[p^\prime\right]\right)_{\tau\sigma^\prime}
.\ee
The matrix element of the projector $P_V$ can be obtained by using the quantum field theoretical framework \cite{Becattini:EPJC52'2007} as
\be\label{M.E.(3)}
\langle p^\prime,\sigma^\prime|P_V|p,\sigma \rangle=\frac{1}{2}\sqrt{\frac{\varepsilon}{\varepsilon^\prime}}\int d^3{\rm x} ~ e^{i\mathbf{x}\cdot(\mathbf{p}-\mathbf{p^\prime})}\left(D^S\left(\left[p^\prime\right]^{-1}\left[p\right]\right)+D^S\left(\left[p^\prime\right]^{\dag}\left[p\right]^{\dag -1}\right)\right)_{\sigma^\prime\sigma}\langle 0|P_V|0 \rangle
,\ee
where $\langle 0|P_V|0 \rangle$ is the vacuum expectation value of the projector $P_V$. For large volume $V$, $\langle 0|P_V|0 \rangle\rightarrow1$, because $P_V$ tends to unity in this case. Substituting eq. \eqref{M.E.(2)} and eq. \eqref{M.E.(3)} in eq. \eqref{M.E.(1)}, we have
\be\label{M.E.(4)}
\nonumber\langle p,\tau|R_{\hat{\omega}}(\phi) P_V|p,\sigma \rangle &=& \int d^3{\rm x} ~ e^{i\mathbf{x}\cdot\left(\mathbf{p}-R_{\hat{\omega}}(\phi)^{-1}(\mathbf{p})\right)} \\ && \times\frac{1}{2}\left(D^S\left(\left[p\right]^{-1}R_{\hat{\omega}}(\phi)\left[p\right]\right)+D^S\left(\left[p\right]^{\dag}R_{\hat{\omega}}(\phi)\left[p\right]^{\dag -1}\right)\right)_{\tau\sigma}
.\ee
Using the unitarity of the Wigner rotation, {\em i.e.}, $$D^S\left(\left[R_{\hat{\omega}}(\phi)(p^\prime)\right]^{-1}R_{\hat{\omega}}(\phi)\left[p^\prime\right]\right)=D^S\left(\left[R_{\hat{\omega}}(\phi)(p^\prime)\right]^{\dag}R_{\hat{\omega}}(\phi)\left[p^\prime\right]^{\dag -1}\right),$$ eq. \eqref{M.E.(4)} can be analytically extended to the imaginary angles. Thus the matrix element in eq. \eqref{M.E.} takes the following form, 
\be\label{M.E.(5)}
\nonumber\langle p,\tau|e^{\beta\omega\hat{j}}P_V|p,\sigma \rangle &=& \int d^3{\rm x} ~ e^{i\mathbf{x}\cdot\left(\mathbf{p}-R_{\hat{\omega}}(i\beta\omega)^{-1}(\mathbf{p})\right)} \\ && \times\frac{1}{2}\left(D^S\left(\left[p\right]^{-1}R_{\hat{\omega}}(i\beta\omega)\left[p\right]\right)+D^S\left(\left[p\right]^{\dag}R_{\hat{\omega}}(i\beta\omega)\left[p\right]^{\dag -1}\right)\right)_{\tau\sigma}
.\ee
It can be observed that, $R_{\hat{\omega}}(i\beta\omega)=e^{\beta\omega J}$. The matrix element 
in eq. \eqref{M.E.} has a spatial integral form, which allows us to write the phase-space 
distribution function \cite{Becattini:AP325'2010} as
\be\label{M.E.(6)}
\nonumber f(x,p)_{\tau\sigma} &=& \left(e^{\beta\left(u_\mu p^\mu+a\right)}+b\right)^{-1}e^{i\mathbf{x}\cdot\left(\mathbf{p}-R_{\hat{\omega}}(i\beta\omega)^{-1}(\mathbf{p})\right)} \\ && \times\frac{1}{2}\left(D^S\left(\left[p\right]^{-1}R_{\hat{\omega}}(i\beta\omega)\left[p\right]\right)+D^S\left(\left[p\right]^{\dag}R_{\hat{\omega}}(i\beta\omega)\left[p\right]^{\dag -1}\right)\right)_{\tau\sigma}
,\ee
where $u_\mu$ accounts for the four-velocity of fluid. For high temperature QCD medium, $\beta\omega$ is small, {\em i.e.} $\beta\omega<1$. Thus the difference between momenta in the exponent of eq. \eqref{M.E.(6)} can be approximated by the lowest order term in $\beta\omega$, 
\be
\nonumber\mathbf{p}-R_{\hat{\omega}}(i\beta\omega)^{-1}(\mathbf{p}) &=& \mathbf{p}-\left[\left(\cosh\beta\omega\right) ~ \mathbf{p}-i\left(\sinh\beta\omega\right) ~ \hat{\omega}\times\mathbf{p}+\left(1-\cosh\beta\omega\right)\mathbf{p} ~ \hat{\omega}\hat{\omega}\right] \\ &\simeq& i\beta ~ \boldsymbol{\omega}\times\mathbf{p}
.\ee
Now, the phase-space distribution function becomes
\be\label{D.F.}
\nonumber f(x,p)_{\tau\sigma} &=& \left(e^{\beta\left(u_\mu p^\mu+a\right)}+b\right)^{-1}e^{-\beta\mathbf{x}\cdot\left(\boldsymbol{\omega}\times\mathbf{p}\right)} \\ && \nonumber\times\frac{1}{2}\left(D^S\left(\left[p\right]^{-1}R_{\hat{\omega}}(i\beta\omega)\left[p\right]\right)+D^S\left(\left[p\right]^{\dag}R_{\hat{\omega}}(i\beta\omega)\left[p\right]^{\dag -1}\right)\right)_{\tau\sigma} \\ &=& \nonumber\left(e^{\beta\left(u_\mu p^\mu+a\right)}+b\right)^{-1}e^{\beta\mathbf{p}\cdot\mathbf{v}} \\ && \times\frac{1}{2}\left(D^S\left(\left[p\right]^{-1}R_{\hat{\omega}}(i\beta\omega)\left[p\right]\right)+D^S\left(\left[p\right]^{\dag}R_{\hat{\omega}}(i\beta\omega)\left[p\right]^{\dag -1}\right)\right)_{\tau\sigma}
,\ee
where $\mathbf{v}=\boldsymbol{\omega}\times\mathbf{x}$ has been used. The above equation is the unnormalized single-particle phase-space distribution function. By taking the trace of the matrix \eqref{D.F.}, one can obtain the phase-space density in $(x,p)$ as
\be\label{D.F.(1)}
\nonumber f(x,p) &=& \sum_{\sigma}f(x,p)_{\sigma\sigma} \\ &=& \left(e^{\beta\left(u_\mu p^\mu+a\right)}+b\right)^{-1}e^{\beta\mathbf{p}\cdot\mathbf{v}}\chi(\beta\omega)
.\ee
In the local rest frame, $\mathbf{v}=0$. So, we have
\be\label{D.F.(2)}
f(x,p)=\frac{1}{e^{\beta\left(u_\mu p^\mu+a\right)}+b}\chi(\beta\omega)
,\ee
where $\chi(\beta\omega)$ represents the angular velocity-dependent part, which is given by
\be\label{A.V.P.}
\nonumber\chi(\beta\omega) &=& Tr\left[D^S\left(R_{\hat{\omega}}(i\beta\omega)\right)\right] \\ &=& \frac{\sinh\left(\left(S+\frac{1}{2}\right)\beta\omega\right)}{\sinh\left(\frac{\beta\omega}{2}\right)}
.\ee
For quarks and antiquarks, spin $S=\frac{1}{2}$ and for gluons, $S=1$. Accordingly, they have different expressions for $\chi(\beta\omega)$ and hence, they have different forms of the distribution 
functions. Thus the quark, antiquark and gluon distribution functions at finite rotation can be written as
\be\label{Q.D.F.}
f_q &=& f^q_0\chi_q(\beta\omega), \\ \label{A.D.F.}\bar{f_q} &=& \bar{f^q_0}\chi_{\bar{q}}(\beta\omega), \\ 
\label{G.D.F.} f_g &=& f^g_0\chi_g(\beta\omega)
,\ee
respectively. The quantities in the above equations are defined as follows, 
\begin{eqnarray}
&&\nonumber f^q_0=\frac{1}{e^{\beta\left(u_\mu p^\mu-\mu\right)}+1}=\frac{1}{e^{\beta\left(\omega_f-\mu\right)}+1}, ~~ \bar{f^q_0}=\frac{1}{e^{\beta\left(u_\mu p^\mu+\mu\right)}+1}=\frac{1}{e^{\beta\left(\omega_f+\mu\right)}+1}, \\
&&\nonumber f^g_0=\frac{1}{e^{\beta u_\mu p^\mu}-1}=\frac{1}{e^{\beta\omega_g}-1}, ~~ \chi_q(\beta\omega)=\frac{\sinh(\beta\omega)}{\sinh(\beta\omega/2)}, \\ 
&&\nonumber \chi_{\bar{q}}(\beta\omega)=\frac{\sinh(\beta\omega)}{\sinh(\beta\omega/2)}, ~~ \chi_g(\beta\omega)=\frac{\sinh(3\beta\omega/2)}{\sinh(\beta\omega/2)} 
,\end{eqnarray}
where $\omega_f$ is the energy of the $f$th flavor of quark (antiquark) and $\omega_g$ is the energy of the gluon in the rotating QGP medium. We note that the flavor degrees of freedom are taken into account 
explicitly. Further, the chemical potentials for flavors, $u$, $d$ and $s$ are kept the same ($\mu_f=\mu$). 

\subsection{Charge conduction in rotating QGP medium}
When a rotating QGP medium is exposed to an external electric field, an electric current gets induced and the corresponding four-current density is written as
\begin{eqnarray}\label{current}
J^\mu = \sum_f g_f \int\frac{d^3\rm{p}}{(2\pi)^3\omega_f}
p^\mu \left[q_f\delta f_q+{\bar q_f}\delta \bar{f_q}\right]
,\end{eqnarray}
where $g_f$, $q_f$ ($\bar q_f$) and $\delta f_q$ ($\delta \bar{f_q}$) represent the degeneracy factor, electric charge and infinitesimal change in the quark (antiquark) distribution function of $f$th 
flavor, respectively. According to Ohm’s law, the spatial component of four-current density is directly proportional to the external electric field with the proportionality factor being the electrical conductivity ($\ec$), {\em i.e.}, 
\begin{eqnarray}\label{Ohm's law}
\mathbf{J}=\ec \mathbf{E}
~.\end{eqnarray}
In order to calculate the infinitesimal change $\delta f_q$, we use the relativistic Boltzmann transport  equation (RBTE) in the novel relaxation time approximation (RTA). In this approximation, the collision 
integral takes a modified form for the validity of the conservation laws \cite{Cercignani:BOOK'2002,Rocha:PRL127'2021}. However, in the frequently used relaxation time approximation, the conservation laws are satisfied, if the relaxation time is momentum-independent. The RBTE in the novel RTA is written as
\be\label{R.B.T.E.1}
\nonumber p^\mu\frac{\partial f_q^\prime}{\partial x^\mu}+q_f F^{\rho\sigma} 
p_\sigma \frac{\partial f_q^\prime}{\partial p^\rho}=-\frac{p_\nu u^\nu}{\tau_f}\left[\delta f_q-\frac{\left\langle\left(\omega_f/\tau_f\right)\delta f_q\right\rangle_0}{\left\langle\omega_f/\tau_f\right\rangle_0}+\frac{P_1^{(0)}\left\langle\left(\omega_f/\tau_f\right)P_1^{(0)}\delta f_q\right\rangle_0}{\left\langle\left(\omega_f/\tau_f\right)P_1^{(0)}P_1^{(0)}\right\rangle_0}\right. \\ \left.+\frac{p^{\left\langle\mu\right\rangle}\left\langle\left(\omega_f/\tau_f\right)p_{\left\langle\mu\right\rangle}\delta f_q\right\rangle_0}{(1/3)\left\langle\left(\omega_f/\tau_f\right)p_{\left\langle\mu\right\rangle}p^{\left\langle\mu\right\rangle}\right\rangle_0}\right]
,\ee
where $\tau_f$ represents the novel relaxation time, $f_q^\prime=\delta f_q+f_q$ with $\delta f_q$ 
denoting the infinitesimal change of the quark distribution function due to the external field and $F^{\rho\sigma}$ is the electromagnetic field strength tensor, whose components are related to the electric and magnetic fields. The novel relaxation time for quark (antiquark), $\tau_f$ ($\tau_{\bar{f}}$) is momentum-dependent with the following form, 
\be\label{N.R.T.}
\tau_{f(\bar{f})}=\left(\beta\omega_f\right)^\xi t_{f(\bar{f})}
,\ee
where $\xi$ is an arbitrary constant that controls the energy dependence of the relaxation time and $t_{f(\bar{f})}$ is momentum-independent, whose form is given \cite{Hosoya:NPB250'1985} by
\begin{eqnarray}
t_{f(\bar{f})}=\frac{1}{5.1T\alpha_s^2\log\left(1/\alpha_s\right)\left[1+0.12(2N_f+1)\right]}
~.\end{eqnarray}
For the calculations, the value of $\xi$ can be taken up to 1, for example, in QCD kinetic theories, 
$\xi=\frac{1}{2}$ \cite{Dusling:PRC81'2010}, while in scalar 
field theories, $\xi=1$ \cite{Calzetta:PRD37'1988}. In eq. \eqref{R.B.T.E.1}, $P_1^{(0)}$ and $p^{\left\langle\mu\right\rangle}$ are respectively defined as
\be\label{E.1}
&&P_1^{(0)}=1-\frac{\left\langle\omega_f/\tau_f\right\rangle_0\omega_f}{\left\langle\omega_f^2/\tau_f\right\rangle_0} , \\ 
&&\label{E.2}p^{\left\langle\mu\right\rangle}=\Delta^{\mu\nu}p_\nu
,\ee
where $\Delta^{\mu\nu}=g^{\mu\nu}-u^\mu u^\nu$. The momentum integral of $A$ ($A$ can be different terms appearing inside the angular brackets in eq. \eqref{R.B.T.E.1}) relative to the local equilibrium 
distribution function is defined as
\be\label{E.3}
\left\langle A \right\rangle_0=\int\frac{d^3\rm{p}}{(2\pi)^3\omega_f}Af_q
~.\ee
Using equations \eqref{E.1}, \eqref{E.2}, \eqref{E.3} and integration by parts, the second (say $L^2$), third (say $L^3$) and fourth (say $L^4$) terms appearing inside the square bracket of eq. \eqref{R.B.T.E.1} can be calculated to take the following forms, 
\be
&&L^2=\delta f_q, \\ 
&&L^3=\frac{\delta f_q\left(1-\frac{\omega_f\int d{\rm p}~\frac{{\rm p}^2}{\tau_f}f_q}{\int d{\rm p}~\frac{{\rm p}^2}{\tau_f}\omega_f f_q}\right)\int d{\rm p}~\frac{{\rm p}^2}{\tau_f}f_q\left(1-\frac{\omega_f\int d{\rm p}~\frac{{\rm p}^2}{\tau_f}f_q}{\int d{\rm p}~\frac{{\rm p}^2}{\tau_f}\omega_f f_q}\right)}{\int d{\rm p}~\frac{{\rm p}^2}{\tau_f}f_q\left(1-\frac{\omega_f\int d{\rm p}~\frac{{\rm p}^2}{\tau_f}f_q}{\int d{\rm p}~\frac{{\rm p}^2}{\tau_f}\omega_f f_q}\right)^2}, \\ 
&&L^4=\frac{3 ~ \delta f_q ~ {\rm p}\int d{\rm p}~\frac{{\rm p}^3}{\tau_f}f_q}{\int d{\rm p}~\frac{{\rm p}^4}{\tau_f}f_q}
.\ee
For the calculation of electrical conductivity, we use the components of $F^{\rho\sigma}$ associated with only electric field. Moreover, in case of a spatially homogeneous distribution function with the steady-state condition, we use $\frac{\partial f_q^\prime}{\partial \mathbf{r}}=0$ and 
$\frac{\partial f_q^\prime}{\partial t}=0$. Thus, RBTE \eqref{R.B.T.E.1} becomes
\be\label{R.B.T.E.2}
q_f\mathbf{E}\cdot\mathbf{p}\frac{\partial f_q^\prime}{\partial p_0}
+q_f p_0\mathbf{E}\cdot\frac{\partial f_q^\prime}{\partial \mathbf{p}}
=-\frac{p_0}{\tau_f}\delta f_q J
,\ee
where the expression of $J$ is written as
\be\label{Q.J}
J=\frac{\left(1-\frac{\omega_f\int d{\rm p}~\frac{{\rm p}^2}{\tau_f}f_q}{\int d{\rm p}~\frac{{\rm p}^2}{\tau_f}\omega_f f_q}\right)\int d{\rm p}~\frac{{\rm p}^2}{\tau_f}f_q\left(1-\frac{\omega_f\int d{\rm p}~\frac{{\rm p}^2}{\tau_f}f_q}{\int d{\rm p}~\frac{{\rm p}^2}{\tau_f}\omega_f f_q}\right)}{\int d{\rm p}~\frac{{\rm p}^2}{\tau_f}f_q\left(1-\frac{\omega_f\int d{\rm p}~\frac{{\rm p}^2}{\tau_f}f_q}{\int d{\rm p}~\frac{{\rm p}^2}{\tau_f}\omega_f f_q}\right)^2}+\frac{3{\rm p}\int d{\rm p}~\frac{{\rm p}^3}{\tau_f}f_q}{\int d{\rm p}~\frac{{\rm p}^4}{\tau_f}f_q}
.\ee
In the antiquark case, $\bar{J}$ is given by
\be\label{A.J}
\bar{J}=\frac{\left(1-\frac{\omega_f\int d{\rm p}~\frac{{\rm p}^2}{\tau_{\bar{f}}}\bar{f_q}}{\int d{\rm p}~\frac{{\rm p}^2}{\tau_{\bar{f}}}\omega_f \bar{f_q}}\right)\int d{\rm p}~\frac{{\rm p}^2}{\tau_{\bar{f}}}\bar{f_q}\left(1-\frac{\omega_f\int d{\rm p}~\frac{{\rm p}^2}{\tau_{\bar{f}}}\bar{f_q}}{\int d{\rm p}~\frac{{\rm p}^2}{\tau_{\bar{f}}}\omega_f \bar{f_q}}\right)}{\int d{\rm p}~\frac{{\rm p}^2}{\tau_{\bar{f}}}\bar{f_q}\left(1-\frac{\omega_f\int d{\rm p}~\frac{{\rm p}^2}{\tau_{\bar{f}}}\bar{f_q}}{\int d{\rm p}~\frac{{\rm p}^2}{\tau_{\bar{f}}}\omega_f \bar{f_q}}\right)^2}+\frac{3{\rm p}\int d{\rm p}~\frac{{\rm p}^3}{\tau_{\bar{f}}}\bar{f_q}}{\int d{\rm p}~\frac{{\rm p}^4}{\tau_{\bar{f}}}\bar{f_q}}
.\ee
Using equations \eqref{Q.D.F.}, \eqref{A.D.F.} and \eqref{N.R.T.} in equations \eqref{Q.J} and \eqref{A.J}, both $J$ and $\bar{J}$ can be simplified, and they are respectively written as
\be\label{Q.J (1)}
&&J=\frac{\left(1-\frac{\omega_f\int d{\rm p}~\frac{{\rm p}^2}{\omega_f^\xi}f^q_0}{\int d{\rm p}~\frac{{\rm p}^2}{\omega_f^{\xi-1}}f^q_0}\right)\int d{\rm p}~\frac{{\rm p}^2}{\omega_f^\xi}f^q_0\left(1-\frac{\omega_f\int d{\rm p}~\frac{{\rm p}^2}{\omega_f^\xi}f^q_0}{\int d{\rm p}~\frac{{\rm p}^2}{\omega_f^{\xi-1}}f^q_0}\right)}{\int d{\rm p}~\frac{{\rm p}^2}{\omega_f^\xi}f^q_0\left(1-\frac{\omega_f\int d{\rm p}~\frac{{\rm p}^2}{\omega_f^\xi}f^q_0}{\int d{\rm p}~\frac{{\rm p}^2}{\omega_f^{\xi-1}}f^q_0}\right)^2}+\frac{3{\rm p}\int d{\rm p}~\frac{{\rm p}^3}{\omega_f^\xi}f^q_0}{\int d{\rm p}~\frac{{\rm p}^4}{\omega_f^\xi}f^q_0}, \\
&&\label{A.J (1)}\bar{J}=\frac{\left(1-\frac{\omega_f\int d{\rm p}~\frac{{\rm p}^2}{\omega_f^\xi}\bar{f^q_0}}{\int d{\rm p}~\frac{{\rm p}^2}{\omega_f^{\xi-1}}\bar{f^q_0}}\right)\int d{\rm p}~\frac{{\rm p}^2}{\omega_f^\xi}\bar{f^q_0}\left(1-\frac{\omega_f\int d{\rm p}~\frac{{\rm p}^2}{\omega_f^\xi}\bar{f^q_0}}{\int d{\rm p}~\frac{{\rm p}^2}{\omega_f^{\xi-1}}\bar{f^q_0}}\right)}{\int d{\rm p}~\frac{{\rm p}^2}{\omega_f^\xi}\bar{f^q_0}\left(1-\frac{\omega_f\int d{\rm p}~\frac{{\rm p}^2}{\omega_f^\xi}\bar{f^q_0}}{\int d{\rm p}~\frac{{\rm p}^2}{\omega_f^{\xi-1}}\bar{f^q_0}}\right)^2}+\frac{3{\rm p}\int d{\rm p}~\frac{{\rm p}^3}{\omega_f^\xi}\bar{f^q_0}}{\int d{\rm p}~\frac{{\rm p}^4}{\omega_f^\xi}\bar{f^q_0}}
.\ee
From eq. \eqref{R.B.T.E.2}, $\delta f_q$ is obtained as
\be
\delta f_q=\frac{2\tau_fq_f\beta}{\omega_fJ}\mathbf{E}\cdot\mathbf{p}f^q_0\left(1-f^q_0\right)\chi_q(\beta\omega)
~.\ee
Similarly, $\delta \bar{f_q}$ for antiquark is determined as
\be
\delta \bar{f_q}=\frac{2\tau_{\bar{f}}{\bar q_f}\beta}{\omega_f\bar{J}}\mathbf{E}\cdot\mathbf{p}\bar{f^q_0}\left(1-\bar{f^q_0}\right)\chi_{\bar{q}}(\beta\omega)
~.\ee
Substituting the values of $\delta f_q$ and $\delta \bar{f_q}$ in eq. (\ref{current}), 
using eq. \eqref{N.R.T.} and then comparing with eq. \eqref{Ohm's law}, we get the 
electrical conductivity for a rotating QGP medium as
\be\label{E.C.}
\sigma_{\rm el} &=& \frac{\beta^{1+\xi}}{3\pi^2}\sum_f g_f q_f^2\int d{\rm p}~\frac{{\rm p}^4}{\omega_f^{2-\xi}}\left[\frac{t_f}{J}f^q_0\left(1-f^q_0\right)\chi_q(\beta\omega)+\frac{t_{\bar{f}}}{\bar{J}}\bar{f^q_0}\left(1-\bar{f^q_0}\right)\chi_{\bar{q}}(\beta\omega)\right]
.\ee

\subsection{Heat conduction in rotating QGP medium}
In a thermal medium, the heat flow four-vector is related to the energy diffusion and the enthalpy diffusion as
\be\label{heat flow (1)}
Q_\mu=\Delta_{\mu\alpha}T^{\alpha\beta}u_\beta-h\Delta_{\mu\alpha}N^\alpha
,\ee
where $T^{\alpha\beta}$ is the energy-momentum tensor, $N^\alpha$ represents the particle flow four-vector, $\Delta_{\mu\alpha}=g_{\mu\alpha}-u_\mu u_\alpha$, the enthalpy per particle $h=(\varepsilon+P)/n$ with $\varepsilon$, $P$ and $n$ denoting the energy density, the pressure and the particle number density, respectively. $N^\alpha$ and $T^{\alpha\beta}$ are also known as the first and the second moments of the distribution function, respectively with the following expressions, 
\be
&&N^\alpha=\sum_f g_f\int \frac{d^3{\rm p}}{(2\pi)^3\omega_f}p^\alpha \left[f_q+\bar{f}_q\right] \label{P.F.F.}, \\ 
&&T^{\alpha\beta}=\sum_f g_f\int \frac{d^3{\rm p}}{(2\pi)^3\omega_f}p^\alpha p^\beta \left[f_q+\bar{f}_q\right] \label{E.M.T.}
.\ee
Using $N^\alpha$ and $T^{\alpha\beta}$, it is possible to get $n=N^\alpha u_\alpha$, $\varepsilon=u_\alpha T^{\alpha\beta} u_\beta$ and $P=-\Delta_{\alpha\beta}T^{\alpha\beta}/3$. In the local rest frame, the heat 
flow is purely spatial, because $Q_\mu u^\mu=0$. Hence, the spatial component of the heat flow four-vector remains finite and is expressed as
\be\label{heat1 (1)}
Q^i=\sum_f g_f\int \frac{d^3{\rm p}}{(2\pi)^3} ~ \frac{p^i}{\omega_f}\left[(\omega_f-h_f)\delta f_q+(\omega_f-\bar{h}_f)\delta \bar{f}_q\right]
.\ee
The Navier-Stokes equation relates the heat flow with the gradients of temperature and pressure as
\be\label{heat.1}
Q^i=-\kappa\delta^{ij}\left[\partial_j T - \frac{T}{\varepsilon+P}\partial_j P\right] 
,\ee
where $\kappa$ is the thermal conductivity. For the calculation of the thermal 
conductivity, the electromagnetic field strength part is not required, so it 
can be dropped from the relativistic Boltzmann transport equation (\ref{R.B.T.E.1}). Now, expanding the gradient of the particle distribution function in terms of the gradients of the flow velocity 
and the temperature in the novel relaxation time approximation, we have 
\be\label{eq1.1}
\nonumber -p^\mu f^q_0\left(1-f^q_0\right)\chi_q(\beta\omega)\left[\left(u_\alpha p^\alpha\right)\partial_\mu\beta+\beta\partial_\mu\left(u_\alpha p^\alpha\right)-\partial_\mu\left(\beta\mu\right)\right]+p^\mu f^q_0\partial_\mu\left(\chi_q(\beta\omega)\right) = -\frac{p_\nu u^\nu}{\tau_f}\left[\delta f_q\right. \\ \left.-\frac{\left\langle\left(\omega_f/\tau_f\right)\delta f_q\right\rangle_0}{\left\langle\omega_f/\tau_f\right\rangle_0}+\frac{P_1^{(0)}\left\langle\left(\omega_f/\tau_f\right)P_1^{(0)}\delta f_q\right\rangle_0}{\left\langle\left(\omega_f/\tau_f\right)P_1^{(0)}P_1^{(0)}\right\rangle_0}+\frac{p^{\left\langle\mu\right\rangle}\left\langle\left(\omega_f/\tau_f\right)p_{\left\langle\mu\right\rangle}\delta f_q\right\rangle_0}{(1/3)\left\langle\left(\omega_f/\tau_f\right)p_{\left\langle\mu\right\rangle}p^{\left\langle\mu\right\rangle}\right\rangle_0}\right]
.\ee
After solving eq. (\ref{eq1.1}), $\delta f_q$ is obtained as
\be\label{delta.q1}
\nonumber\delta f_q &=& -\frac{\beta\tau_ff^q_0\left(1-f^q_0\right)}{J}\chi_q(\beta\omega)\left[\frac{\left(\omega_f-h_f\right)p^j}{T\omega_f}\left(\partial_jT-\frac{T}{\varepsilon+P}\partial_jP\right)+p_0\frac{DT}{T}-\frac{p^\mu p^\alpha}{p_0}\nabla_\mu u_\alpha\right. \\ && \left.+TD\left(\frac{\mu}{T}\right)\right]-\frac{\beta\tau_ff^q_0}{J}\sinh\left(\frac{\beta\omega}{2}\right)\left[TD\left(\frac{\omega}{T}\right)+\frac{Tp^\alpha}{p_0}\nabla_\alpha\left(\frac{\omega}{T}\right)\right]
.\ee
In the similar way, $\delta \bar{f}_q$ for antiquark is determined as
\be\label{delta.aq1}
\nonumber\delta \bar{f_q} &=& -\frac{\beta\tau_{\bar{f}}\bar{f^q_0}\left(1-\bar{f^q_0}\right)}{\bar{J}}\chi_{\bar{q}}(\beta\omega)\left[\frac{\left(\omega_f-\bar{h}_f\right)p^j}{T\omega_f}\left(\partial_jT-\frac{T}{\varepsilon+P}\partial_jP\right)+p_0\frac{DT}{T}-\frac{p^\mu p^\alpha}{p_0}\nabla_\mu u_\alpha\right. \\ && \left.-TD\left(\frac{\mu}{T}\right)\right]-\frac{\beta\tau_{\bar{f}}\bar{f^q_0}}{\bar{J}}\sinh\left(\frac{\beta\omega}{2}\right)\left[TD\left(\frac{\omega}{T}\right)+\frac{Tp^\alpha}{p_0}\nabla_\alpha\left(\frac{\omega}{T}\right)\right]
.\ee
Substituting the values of $\delta f_q$ and $\delta \bar{f_q}$ in eq. (\ref{heat1 (1)}), 
using eq. \eqref{N.R.T.} and then comparing with eq. (\ref{heat.1}), we get the thermal 
conductivity for a rotating QGP medium as
\be\label{T.C.}
\nonumber\kappa &=& \frac{\beta^{2+\xi}}{6\pi^2}\sum_fg_f\int d{\rm p}~\frac{{\rm p}^4}{\omega_f^{2-\xi}}\left[\frac{t_f}{J}(\omega_f-h_f)^2f^q_0\left(1-f^q_0\right)\chi_q(\beta\omega)\right. \\ && \left.+\frac{t_{\bar{f}}}{\bar{J}}(\omega_f-\bar{h}_f)^2\bar{f^q_0}\left(1-\bar{f^q_0}\right)\chi_{\bar{q}}(\beta\omega)\right]
.\ee

\subsection{Observables in rotating QGP medium}
The observables such as the Knudsen number, the elliptic flow, the specific heat at constant pressure, the specific heat at constant volume, the trace anomaly, the thermal diffusion constant and the isothermal compressibility are also likely to be affected by the rotation of the QGP medium. 

The Knudsen number is the ratio of the mean free path ($\lambda$) to the 
characteristic length scale ($l$) of the medium, {\em i.e.} $\Omega={\lambda}/{l}$. 
For the medium to be in the equilibrium state, $l$ should be larger than $\lambda$. 
Using $\lambda={3\kappa}/{(vC_V)}$, the Knudsen number becomes
\be\label{Knudsen number}
\Omega=\frac{3\kappa}{lvC_V}
~,\ee
where $v$ and $C_V$ are the relative speed and the specific heat at constant 
volume, respectively. The Knudsen number explains the competition 
between the microscopic and macroscopic length scales of the concerned system. 
If $\Omega$ is less than unity, then the equilibrium hydrodynamics can be 
applicable and in the hydrodynamic limit, the system assumes local thermodynamical 
equilibrium. On the other hand, the Knudsen number is large for rarefied 
gases and the regime of extremely large $\Omega$ is known as the free streaming 
particle regime. In heavy ion collisions, the degree of thermalization in the 
produced matter can be characterized by the Knudsen number. The experimental 
extraction of the Knudsen number may rely on various quantities, such as the 
overlap area between two colliding nuclei in the transverse plane of the 
collision zone, eccentricity, multiplicity density, centrality etc. It is reported 
in ref. \cite{Bhalerao:PLB627'2005} that the inverse of the Knudsen number 
can be obtained experimentally as it is directly proportional to the multiplicity 
density and inversely proportional to the transverse area of the collision zone. In this work, we have 
used $v\simeq 1$ and $l=1$ fm, and calculated $C_V$ from the energy-momentum tensor as follows, 
\be\label{S.V.}
\nonumber C_V &=& \frac{\partial (u_\mu T^{\mu\nu}u_\nu)}{\partial T} \\ &=& \nonumber\sum_f g_f\int \frac{d^3{\rm p}}{(2\pi)^3} ~ \omega_f\left[\frac{\partial f_q}{\partial T}+\frac{\partial\bar{f}_q}{\partial T}\right]+g_g\int \frac{d^3{\rm p}}{(2\pi)^3} ~ \omega_g\frac{\partial f_g}{\partial T} \\ &=& \nonumber\frac{1}{2\pi^2}\sum_fg_f\int d{\rm p} ~ {\rm p}^2\omega_f\left[\chi_q(\beta\omega)\frac{\partial f^q_0}{\partial T}+f^q_0\frac{\partial \chi_q(\beta\omega)}{\partial T}+\chi_{\bar{q}}(\beta\omega)\frac{\partial \bar{f^q_0}}{\partial T}+\bar{f^q_0}\frac{\partial \chi_{\bar{q}}(\beta\omega)}{\partial T}\right] \\ && +\frac{1}{2\pi^2}g_g\int d{\rm p} ~ {\rm p}^2\omega_g\left[\chi_g(\beta\omega)\frac{\partial f^g_0}{\partial T}+f^g_0\frac{\partial \chi_g(\beta\omega)}{\partial T}\right]
.\ee
In the above equation, the values of different partial derivatives are given by
\be
&&\frac{\partial f^q_0}{\partial T}=\beta^2(\omega_f-\mu)f^q_0\left(1-f^q_0\right), \\
&&\frac{\partial \bar{f^q_0}}{\partial T}=\beta^2(\omega_f+\mu)\bar{f^q_0}\left(1-\bar{f^q_0}\right), \\
&&\frac{\partial f^g_0}{\partial T}=\beta^2\omega_gf^g_0\left(1+f^g_0\right), \\
&&\frac{\partial \chi_q(\beta\omega)}{\partial T}=\frac{\partial \chi_{\bar{q}}(\beta\omega)}{\partial T}=-\beta^2\omega\sinh\left(\frac{\beta\omega}{2}\right), \\
&&\frac{\partial \chi_g(\beta\omega)}{\partial T}=- 2\beta^2\omega\sinh(\beta\omega)
.\ee
Now, eq. \eqref{S.V.} becomes
\be\label{S.V.(1)}
\nonumber C_V &=& \frac{\beta^2}{2\pi^2}\sum_fg_f\int d{\rm p} ~ {\rm p}^2\omega_f\left[\left\lbrace\left(\omega_f-\mu\right)f^q_0\left(1-f^q_0\right)+\left(\omega_f+\mu\right)\bar{f^q_0}\left(1-\bar{f^q_0}\right)\right\rbrace\frac{\sinh(\beta\omega)}{\sinh(\beta\omega/2)}\right. \\ && \left.\nonumber -\left(f^q_0+\bar{f^q_0}\right)\omega\sinh(\beta\omega/2)\right]+\frac{\beta^2}{2\pi^2}g_g\int d{\rm p} ~ {\rm p}^2\omega_g\left[\omega_gf^g_0\left(1+f^g_0\right)\frac{\sinh(3\beta\omega/2)}{\sinh(\beta\omega/2)}\right. \\ && \left.- 2f^g_0\omega\sinh(\beta\omega)\right]
.\ee
Using the values of $\kappa$ \eqref{T.C.} and $C_V$ \eqref{S.V.(1)} 
in eq. \eqref{Knudsen number}, the Knudsen number for the QGP 
medium at finite rotation can be explored. 

The elliptic flow ($v_2$) is one of the key observables and represents the azimuthal anisotropy in the production of particles in ultrarelativistic heavy ion collisions at terrestrial laboratories. In such collisions, the overlapping region of bombarding nuclei is the origin of generation of the elliptic 
flow, where the pressure gradient converts the initial spatial 
anisotropy (quantified by the eccentricity $\epsilon$) into the momentum anisotropy. The 
elliptic flow develops gradually with the evolution of the system. The anisotropy of the momentum 
distributions can be completely attained only once all parts of the system get involved in the 
process. It also gives the information about the degree of interactions/reinteractions among the 
particles produced in heavy ion collisions. Thus, its magnitude is associated with the equilibrium 
property of the thermal medium and any deviation from equilibrium can reduce the magnitude of such 
flow. Experimentally, elliptic flow can be analyzed by selecting events in a centrality class. The 
quantities which influence the elliptic flow, such as the density, the transverse size and the 
eccentricity, fluctuate from one event to the other, thus causing dynamical fluctuations of the 
elliptic flow. The elliptic flow is closely related to the Knudsen number and is defined \cite{Bhalerao:PLB627'2005,Drescher:PRC76'2007,Gombeaud:PRC77'2008} as
\be\label{Elliptic flow}
v_2=\frac{v_2^h}{1+\frac{\Omega}{\Omega_h}}
.\ee
Here, $v_2^h$ denotes the elliptic flow in the hydrodynamic limit ($\Omega\rightarrow 0$ limit) and $\Omega_h$ is the value of the Knudsen number calculated by observing the transition between the hydrodynamic regime and the free streaming particle regime. The transport calculation \cite{Gombeaud:PRC77'2008} has estimated $\Omega_h \approx 0.7$ and $v_2^h \approx 0.1$. The presence of magnetic field, temperature and chemical potential was observed to influence the elliptic flow \cite{Mohapatra:MPLA26'2011,Rath:EPJA59'2023,Rath:EPJC83'2023}. Similarly, the finite rotation could also affect the elliptic flow. Equation \eqref{Elliptic flow} is a general relation between the Knudsen number and the elliptic flow as long as the system is not far away from the hydrodynamic limit, where the mean free path is less than the characteristic length scale of the medium \cite{Bhalerao:PLB627'2005,Drescher:PRC76'2007,Gombeaud:PRC77'2008}. However, this relation may not be valid, if the medium is far away from the equilibrium. This relation is defined within the applicability of the Boltzmann transport theory and is appropriate to our present study where we assume that the medium stays near the equilibrium with an infinitesimal deviation from the equilibrium. Additionally, eq. \eqref{Elliptic flow} can describe the equation of state, the equilibrium property of the medium, the transition between the hydrodynamic regime and the free-streaming particle regime, the thermalization of the medium, the partonic cross sections, the centrality dependence and the system-size dependence of the elliptic flow as well as the effects of various extreme conditions on the elliptic flow through the Knudsen number. The extreme conditions include the strong magnetic fields, high temperatures, high densities and rapid rotations. Works on the effects of magnetic field, temperature and density on the elliptic flow using the aforesaid relation have already been done \cite{Rath:EPJA59'2023,Rath:EPJC83'2023,Shaikh:PRD108'2023}. Out of the aforesaid extreme conditions, the effect of rapid rotation on the elliptic flow has been studied in the present work. The extreme conditions do not modify the form of equation relating the Knudsen number and elliptic flow, rather their effects conspicuously alter the Knudsen number, thus influencing the elliptic flow. We have used eq. \eqref{Elliptic flow} to explore the elliptic flow for the rotating medium. Substituting the value of the Knudsen number \eqref{Knudsen number} in eq. \eqref{Elliptic flow}, the effect of finite rotation on the elliptic flow can be investigated. 

Like the specific heat at constant volume, it is possible to determine the specific heat at constant pressure from the energy-momentum tensor as
\be\label{S.P.}
\nonumber C_P &=& \frac{\partial (u_\mu T^{\mu\nu}u_\nu-\Delta_{\mu\nu}T^{\mu\nu}/3)}{\partial T} \\ &=& \nonumber\sum_f g_f\int \frac{d^3{\rm p}}{(2\pi)^3} ~ \omega_f\left[\frac{\partial f_q}{\partial T}+\frac{\partial\bar{f}_q}{\partial T}\right]+\frac{1}{3}\sum_f g_f\int \frac{d^3{\rm p}}{(2\pi)^3} ~ \frac{{\rm p}^2}{\omega_f}\left[\frac{\partial f_q}{\partial T}+\frac{\partial\bar{f}_q}{\partial T}\right] \\ && \nonumber+g_g\int \frac{d^3{\rm p}}{(2\pi)^3} ~ \omega_g\frac{\partial f_g}{\partial T}+\frac{1}{3}g_g\int \frac{d^3{\rm p}}{(2\pi)^3} ~ \frac{{\rm p}^2}{\omega_g}\frac{\partial f_g}{\partial T} \\ &=& \nonumber\frac{1}{2\pi^2}\sum_fg_f\int d{\rm p} ~ {\rm p}^2\omega_f\left[\chi_q(\beta\omega)\frac{\partial f^q_0}{\partial T}+f^q_0\frac{\partial \chi_q(\beta\omega)}{\partial T}+\chi_{\bar{q}}(\beta\omega)\frac{\partial \bar{f^q_0}}{\partial T}+\bar{f^q_0}\frac{\partial \chi_{\bar{q}}(\beta\omega)}{\partial T}\right] \\ && \nonumber+\frac{1}{6\pi^2}\sum_fg_f\int d{\rm p} ~ \frac{{\rm p}^4}{\omega_f}\left[\chi_q(\beta\omega)\frac{\partial f^q_0}{\partial T}+f^q_0\frac{\partial \chi_q(\beta\omega)}{\partial T}+\chi_{\bar{q}}(\beta\omega)\frac{\partial \bar{f^q_0}}{\partial T}+\bar{f^q_0}\frac{\partial \chi_{\bar{q}}(\beta\omega)}{\partial T}\right] \\ && \nonumber+\frac{1}{2\pi^2}g_g\int d{\rm p} ~ {\rm p}^2\omega_g\left[\chi_g(\beta\omega)\frac{\partial f^g_0}{\partial T}+f^g_0\frac{\partial \chi_g(\beta\omega)}{\partial T}\right] \\ && +\frac{1}{6\pi^2}g_g\int d{\rm p} ~ \frac{{\rm p}^4}{\omega_g}\left[\chi_g(\beta\omega)\frac{\partial f^g_0}{\partial T}+f^g_0\frac{\partial \chi_g(\beta\omega)}{\partial T}\right]
.\ee
Substituting the values of partial derivatives in the above equation and simplifying, we get 
\be\label{S.P.(1)}
\nonumber C_P &=& \frac{\beta^2}{6\pi^2}\sum_fg_f\int d{\rm p}\left(3{\rm p}^2\omega_f+\frac{{\rm p}^4}{\omega_f}\right)\left[\left\lbrace\left(\omega_f-\mu\right)f^q_0\left(1-f^q_0\right)+\left(\omega_f+\mu\right)\bar{f^q_0}\left(1-\bar{f^q_0}\right)\right\rbrace\right. \\ && \left.\nonumber\times\frac{\sinh(\beta\omega)}{\sinh(\beta\omega/2)}-\left(f^q_0+\bar{f^q_0}\right)\omega\sinh(\beta\omega/2)\right] \\ && +\frac{\beta^2}{6\pi^2}g_g\int d{\rm p}\left(3{\rm p}^2\omega_g+\frac{{\rm p}^4}{\omega_g}\right)\left[\omega_gf^g_0\left(1+f^g_0\right)\frac{\sinh(3\beta\omega/2)}{\sinh(\beta\omega/2)}-2f^g_0\omega\sinh(\beta\omega)\right]
.\ee
From $C_P$ \eqref{S.P.(1)} and $C_V$ \eqref{S.V.(1)}, the adiabatic index can be determined and studied to understand the effect of rotation on it, which is defined as
\be\label{Ratio}
\gamma=\frac{C_P}{C_V}
.\ee
This ratio measures how much internal energy is getting converted into work in any compression/expansion process. Different systems possess different $\gamma$ values, {\em e.g.} for a nonrelativistic ideal 
gas, $\gamma=\frac{5}{3}$ and for a system with massless particles, $\gamma=\frac{4}{3}$. The adiabatic 
index for an ideal gas is also associated with the degrees of freedom, thus, for monatomic, diatomic 
and triatomic gases, the values of $\gamma$ are different. As $C_P$ and $C_V$ for the system of quarks, antiquarks and gluons are influenced by the extreme conditions, their ratio is supposed to change in 
such conditions. Thus the adiabatic index depends on the features of the medium and its physical 
conditions. Since the rotation influences the properties of the QGP medium, the value of the adiabatic index must be affected. 

The trace anomaly is also one of the important thermodynamic observables, that characterizes the deviation 
of the system from the conformal limit, where the energy density ($\varepsilon$) equals 3 times the 
pressure ($3P$). In the ideal case or in the noninteracting case, {\em i.e.} when the quarks and gluons 
are idealized to be noninteracting, massless and there is no net baryon number in the system, the trace 
anomaly vanishes and the conformal symmetry is approached. On the other hand, if the trace anomaly is 
finite, it describes the existence of interactions among partons and the generation of mass. It has been observed that the trace anomaly could approach zero for large values of the temperature in QCD due to the asymptotic freedom \cite{Andersen:PRD93'2016}. The study on the variation of the trace anomaly 
with the temperature is also very important in determining the phase transition temperature. In this work, 
we have explored how the trace anomaly gets altered by the rotation of the QGP medium. The trace anomaly 
or interaction measure is defined as
\be\label{T.A.}
\nonumber \Gamma &=& \varepsilon-3P \\ &=& \nonumber u_\mu T^{\mu\nu}u_\nu+\Delta_{\mu\nu}T^{\mu\nu} \\ &=& \nonumber\frac{1}{2\pi^2}\sum_fg_f\int d{\rm p}\left({\rm p}^2\omega_f-\frac{{\rm p}^4}{\omega_f}\right)\left[f^q_0\chi_q(\beta\omega)+\bar{f^q_0}\chi_{\bar{q}}(\beta\omega)\right] \\ && +\frac{1}{2\pi^2}g_g\int d{\rm p}\left({\rm p}^2\omega_g-\frac{{\rm p}^4}{\omega_g}\right)f^g_0\chi_g(\beta\omega)
.\ee

The thermal diffusion constant ($D^T$) is associated with the rate of the heat transfer in a system. It appears 
in the diffusion equation for heat as
\be\label{H.E.}
\frac{\partial T}{\partial t}=D^T\nabla^2T
,\ee
which is carried out at constant pressure. The thermal diffusion constant in the first order relativistic viscous hydrodynamics can be defined in terms of the thermal conductivity and the specific heat at 
constant pressure as
\be\label{T.D.C.}
D^T=\frac{\kappa}{C_P}
.\ee
Larger value of the thermal diffusion constant reflects the fact that the heat transfer is faster in the medium. Through $\kappa$ \eqref{T.C.} and $C_P$ \eqref{S.P.(1)}, it is possible to discern the influence of rotation on the thermal diffusion constant in QGP medium. 

Another important observable is the isothermal compressibility ($k^T$), which can also be modulated by the rotation of the QGP medium. It acts as one of the thermodynamic response functions related to the equation 
of state, which governs the evolution of the system. It can also be useful in discerning the nature of the phase transition of the matter produced in heavy ion collisions. Depending on the center-of-mass energy, $k^T$ may also change. Thermodynamically, the isothermal compressibility is defined as
\be\label{I.C.}
k^T=-\frac{1}{V}\left(\frac{\partial V}{\partial P}\right)_T
,\ee
where $V$ denotes the volume. Thus the isothermal compressibility measures how much the volume of the 
system decreases with the increase of pressure at a constant temperature. Generally, an increase 
in the applied pressure leads to a decrease in volume, so the negative sign appearing 
in eq. \eqref{I.C.} makes the value of $k^T$ positive. Smaller value of the isothermal compressibility  indicates that the matter is stiffer. This observable can also provide the information about the 
deviation of a system from perfect fluid behavior. In heavy ion collisions, the isothermal 
compressibility can be calculated from the event-by-event fluctuation of charged particle multiplicity distributions \cite{Mrowczynski:PLB430'1998}. For precise idea about the equation of state of the 
system, $k^T$ has utmost significance. For calculation, the isothermal compressibility can 
be transformed into a tractable form as
\be\label{I.C.(1)}
\nonumber k^T &=& \frac{1}{n}\left(\frac{\partial n}{\partial P}\right)_T \\ &=& \frac{1}{n}\left(\frac{\partial n/\partial \mu}{\partial P/\partial \mu}\right)_T
.\ee
Here, the expressions of $n$, $\partial n/\partial \mu$ and $\partial P/\partial \mu$ are 
respectively written as
\be
&&n=\frac{1}{2\pi^2}\sum_fg_f\int d{\rm p} ~ {\rm p}^2\left[f^q_0\chi_q(\beta\omega)+\bar{f^q_0}\chi_{\bar{q}}(\beta\omega)\right]+\frac{1}{2\pi^2}g_g\int d{\rm p} ~ {\rm p}^2f^g_0\chi_g(\beta\omega), \\
&&\frac{\partial n}{\partial \mu}=\frac{\beta}{2\pi^2}\sum_fg_f\int d{\rm p} ~ {\rm p}^2\left[f^q_0\left(1-f^q_0\right)\chi_q(\beta\omega)-\bar{f^q_0}\left(1-\bar{f^q_0}\right)\chi_{\bar{q}}(\beta\omega)\right], \\
&&\frac{\partial P}{\partial \mu}=\frac{\beta}{6\pi^2}\sum_fg_f\int d{\rm p} ~ \frac{{\rm p}^4}{\omega_f}\left[f^q_0\left(1-f^q_0\right)\chi_q(\beta\omega)-\bar{f^q_0}\left(1-\bar{f^q_0}\right)\chi_{\bar{q}}(\beta\omega)\right] 
.\ee
By using the values of $n$, $\partial n/\partial \mu$ and $\partial P/\partial \mu$ in eq. \eqref{I.C.(1)}, the isothermal compressibility at finite rotation can be obtained. 

\section{Results and discussions}
In the present work, the quasiparticle masses or the thermal masses of particles within the quasiparticle model are considered. These masses are generated due to the interactions of partons with the 
surrounding thermal medium. At finite temperature and finite chemical potential, the quasiparticle masses (squared) of quark and gluon up to one-loop are given \cite{Braaten:PRD45'1992,Peshier:PRD66'2002,Blaizot:PRD72'2005} by
\be\label{Q.P.M.(Quark mass)}
&& m_{fT}^2=\frac{g^2T^2}{6}\left(1+\frac{\mu_f^2}{\pi^2T^2}\right), \\
&&\label{Q.P.M.(Gluon mass)}m_{gT}^2=\frac{g^2T^2}{6}\left(N_c+\frac{N_f}{2}+\frac{3}{2\pi^2T^2}\sum_f\mu_f^2\right)
,\ee
respectively. Here, $g$ is the one-loop running coupling with the following \cite{Kapusta:BOOK'2006} form, 
\begin{eqnarray}
g^2=4\pi\alpha_s=\frac{48\pi^2}{\left(11N_c-2N_f\right)\ln\left({\Lambda^2}/{\Lambda_{\rm\overline{MS}}^2}\right)}
~,\end{eqnarray}
where $\Lambda_{\rm\overline{MS}}=0.176$ GeV, $\Lambda=2\pi\sqrt{T^2+\mu_f^2/\pi^2}$ for electrically charged particles (quarks and antiquarks) and $\Lambda=2 \pi T$ for gluons. In this work, the value of the chemical potential is set at 0.06 GeV. 

\begin{figure}[]
\begin{center}
\begin{tabular}{c c}
\includegraphics[width=7.4cm]{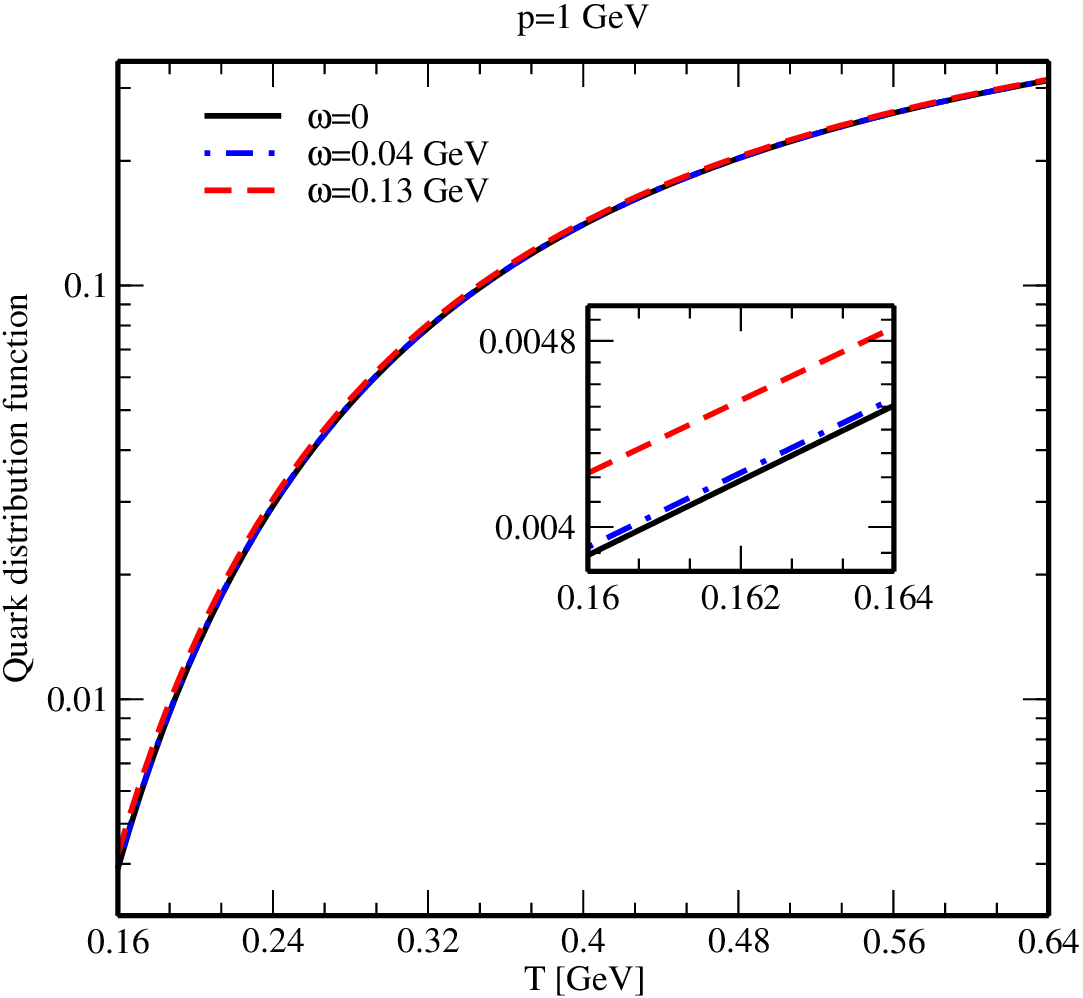}&
\hspace{0.73 cm}
\includegraphics[width=7.4cm]{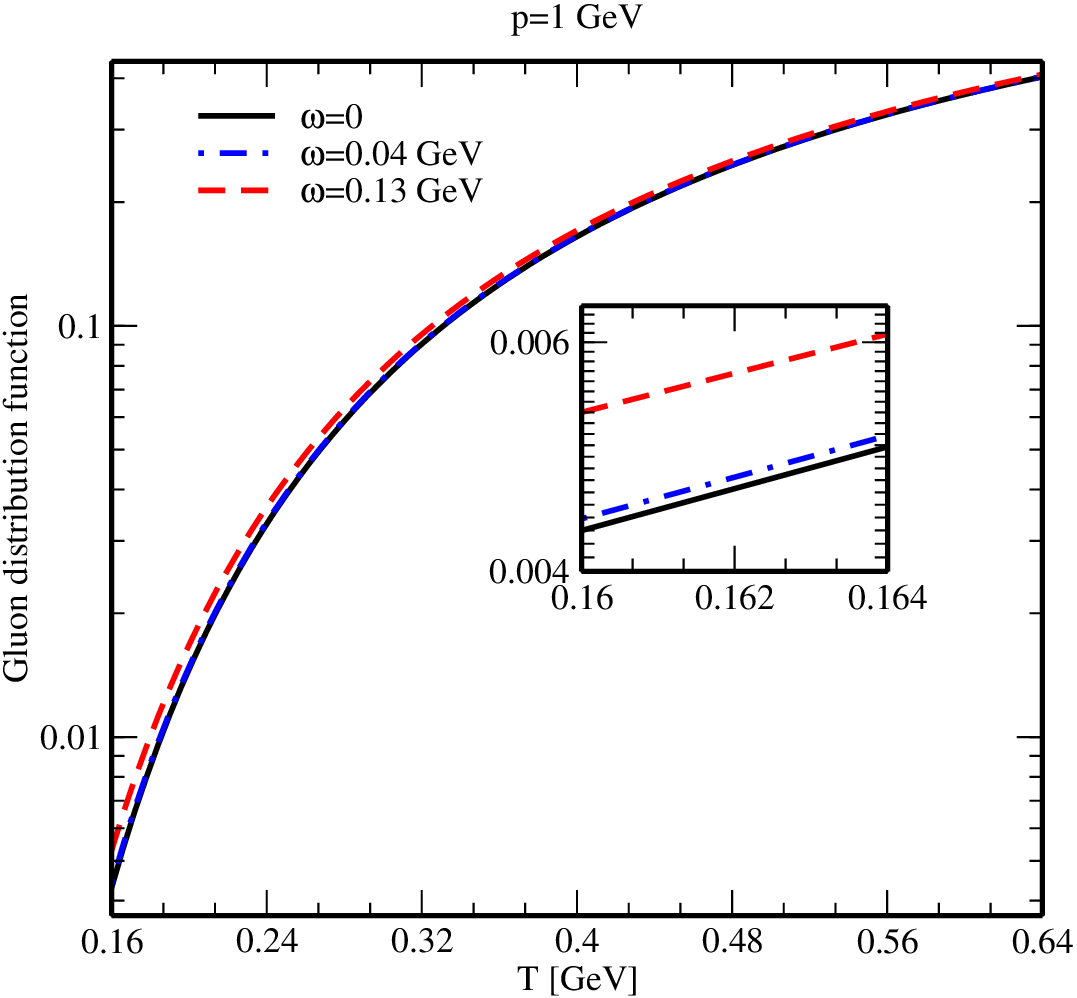} \\
a & b
\end{tabular}
\caption{Temperature dependence of (a) quark and (b) gluon distribution functions for different values of angular velocity.}\label{Fig.tdf}
\end{center}
\end{figure}

\begin{figure}[]
\begin{center}
\begin{tabular}{c c}
\includegraphics[width=7.4cm]{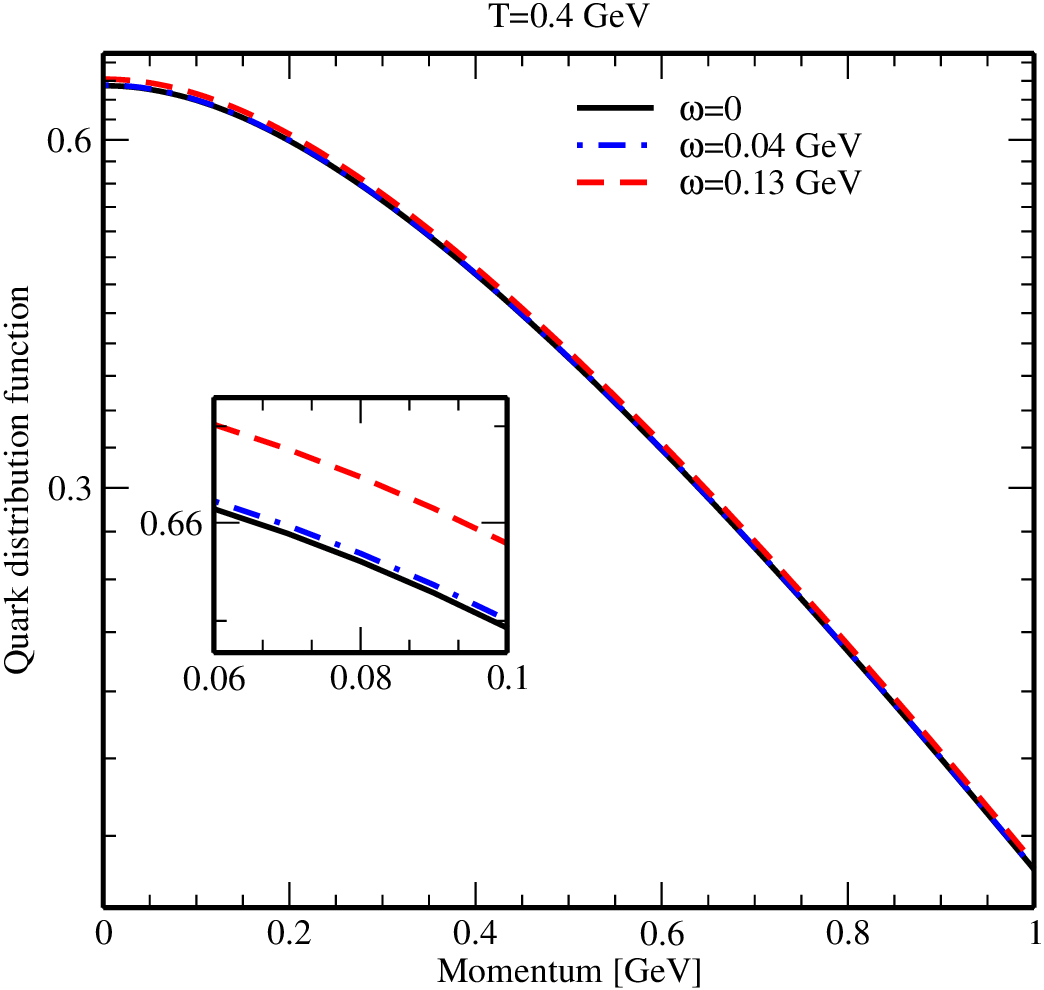}&
\hspace{0.73 cm}
\includegraphics[width=7.4cm]{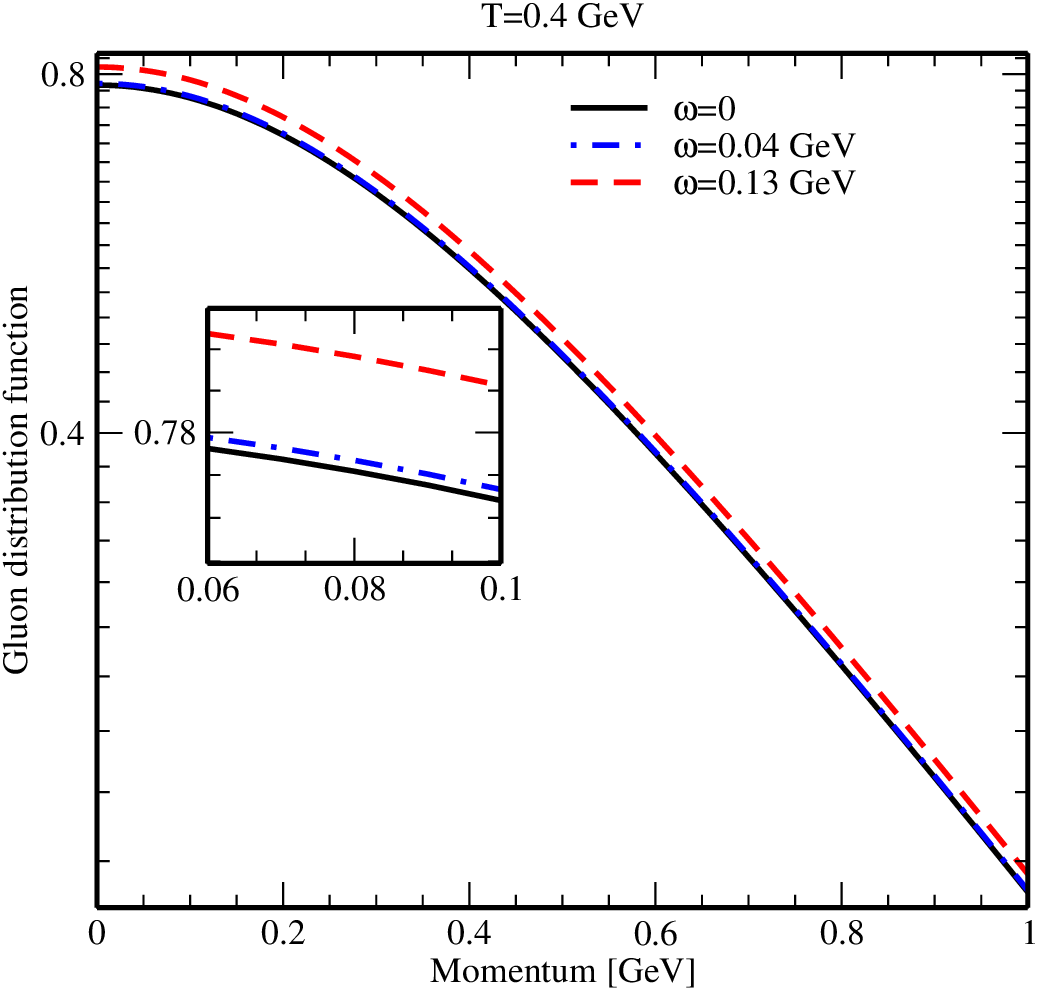} \\
a & b
\end{tabular}
\caption{Momentum dependence of (a) quark and (b) gluon distribution functions for different values of angular velocity.}\label{Fig.pdf}
\end{center}
\end{figure}

To understand the effect of rotational QGP medium on the transport coefficients and observables, let us first observe how the parton distribution functions get altered in the similar environment. We note that, 
in the nonrotating limit ($\omega\to 0$), the $\omega$-dependent factor reduces to $\chi(\beta\omega)\to 2S+1$, which corresponds to the spin degeneracy factor. Therefore, we include the spin degeneracy factor in the nonrotating limit for comparison of the parton distribution functions between the rotating and the nonrotating scenarios at the equal base. Figure \ref{Fig.tdf} shows the temperature dependence of the $u$ quark and gluon distribution functions at a fixed momentum for different values of the angular velocity, whereas figure \ref{Fig.pdf} displays the momentum dependence of these distribution functions at a fixed temperature for different angular velocities. From the observation on the distribution functions, it is 
inferred that the parton number densities increase with the increase of angular velocity. Further, it is found from figures \ref{Fig.tdf} and \ref{Fig.pdf} that the changes in the parton distribution functions due to rotation are modest when changing the value of angular velocity ($\omega$) from 0 to $0.04$ GeV and from $0.04$ GeV to $0.13$ GeV. Quantitatively, this can be comprehended from the $\omega$-dependent factors appearing in the parton distribution functions. The shift of the $\omega$-dependent factor containing sine hyperbolic functions is very small for $\omega=0.04$ GeV as compared to the nonrotating scenario. Similarly, the change of the $\omega$-dependent factor is small when changing the value of $\omega$ from 0.04 GeV to 0.13 GeV, thus indicating a marginal shift of the distribution function. Qualitatively, this can be understood from the fact that, for a rotating medium with low angular velocity, the temperature acts as the dominant energy scale, thus angular velocity leaves deficient impact on the phase-space distribution functions when they are plotted as functions of the temperature, unlike the adequate impact in the high angular velocity regime. It is further noticed from figures \ref{Fig.tdf} and \ref{Fig.pdf} that, as compared to the quark distribution function, the gluon distribution function is more influenced by the rotation of the QGP medium. This can be perceived from the fact that the number of gluons in adjoint representation of SU(3) is larger than the number of quarks in fundamental representation of SU(3), so the overall response of gluons to rotation exceeds that of 
quarks. These observations on the parton distribution functions would help greatly in exploring the effect of rotation on various transport coefficients and observables in kinetic theory approach. 

\begin{figure}[]
\begin{center}
\begin{tabular}{c c}
\includegraphics[width=7.4cm]{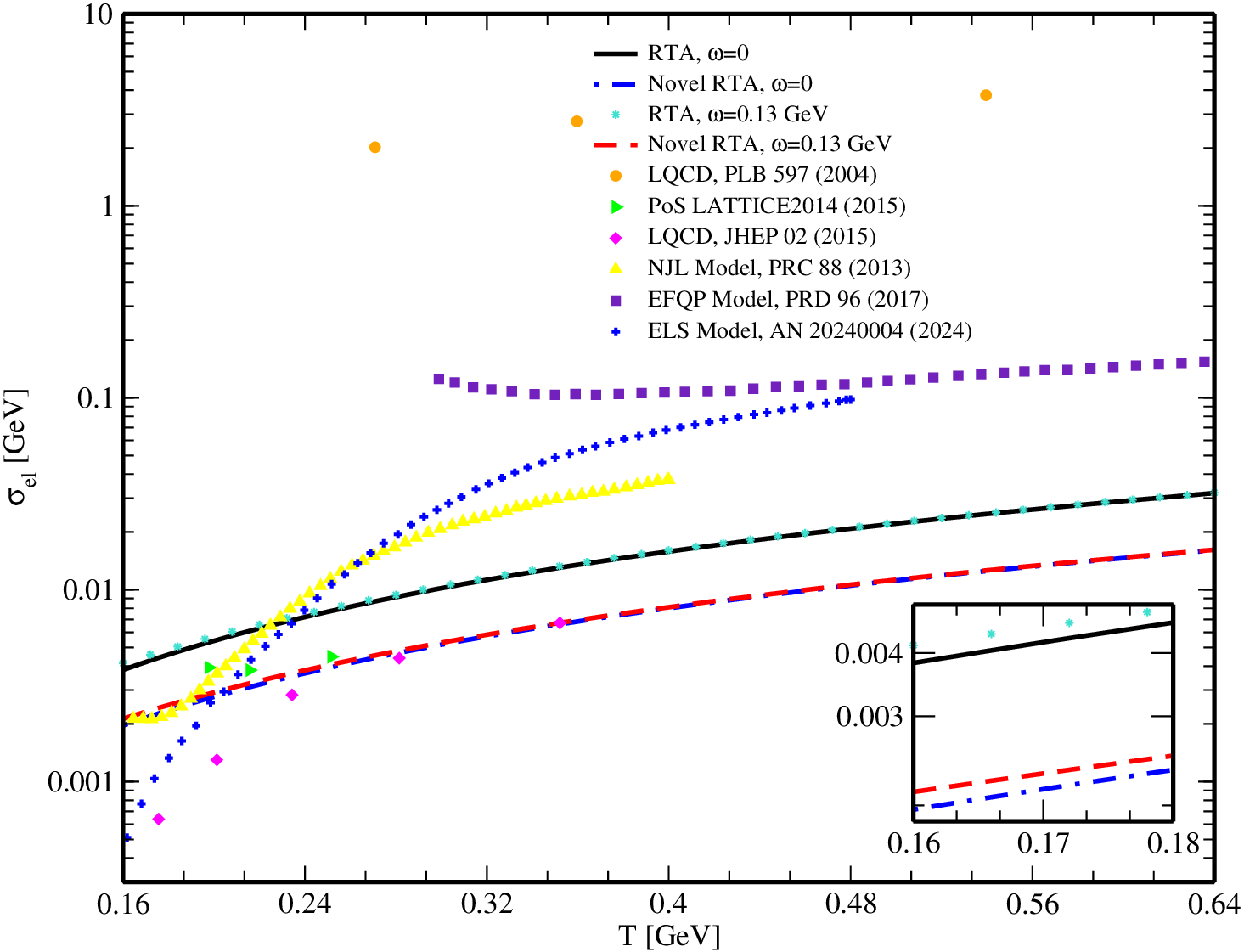}&
\hspace{0.73 cm}
\includegraphics[width=7.4cm]{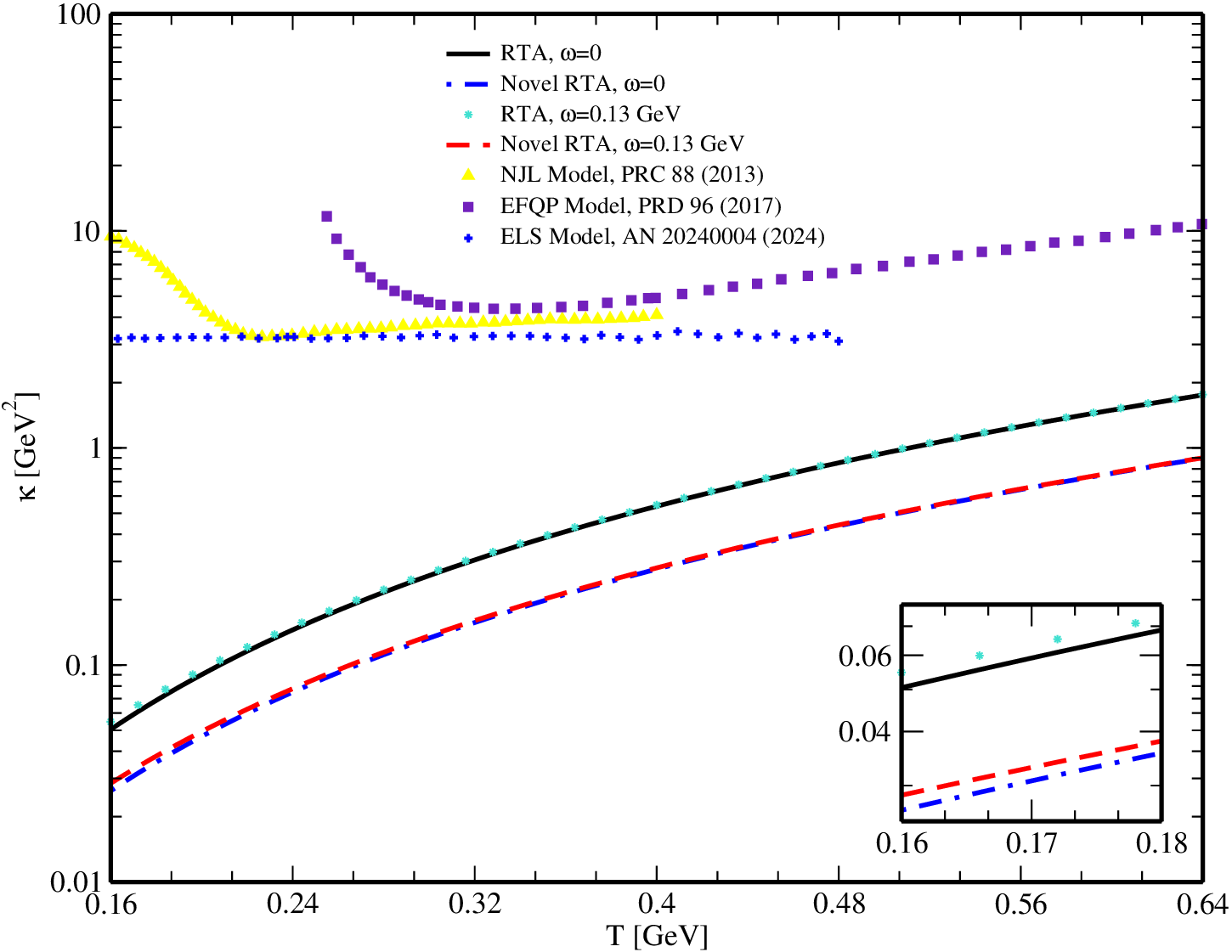} \\
a & b
\end{tabular}
\caption{Variations of (a) the electrical conductivity and (b) the thermal conductivity with temperature for different values of angular velocity.}\label{Fig.1}
\end{center}
\end{figure}

Figure \ref{Fig.1} depicts the variations of the electrical ($\ec$) and the thermal ($\kappa$) conductivities 
of the QGP medium with temperature for different values of angular velocity 
in the relaxation time approximation as well as in the novel relaxation time approximation. Our observation shows that the emergence of angular velocity gives an increasing effect to the electrical conductivity and the thermal conductivity. This increase in $\ec$ and $\kappa$ is mainly due to the increase in the distribution functions at finite angular velocity. The effect of $\omega$ on the conductivities is as modest as on the distribution functions, which can be perceived from the fact that, in the nonrotating limit, the $\omega$-dependent factor reduces to $\chi(\beta\omega)\rightarrow 2S+1$, which corresponds to the spin degeneracy factor. Thus, the spin degeneracy factor is included within the total degeneracy factor for $\omega=0$ scenario and we divide the total degeneracy factor (that already contains the spin degeneracy factor) by the aforementioned spin degeneracy factor to avoid the double counting of the spin degeneracy factor for $\omega\neq 0$ scenario used in the comparison for various transport coefficients and observables. This contributes to the meagre effect of rotation on the abovementioned conductivities. Additionally, the presence of other factors ($J$, $\bar{J}$ etc.) in the expressions of $\ec$ and $\kappa$ suppresses the increase due to rotation, thus leaving a smaller impact of $\omega$ on conductivities. It is also observed that the effect of angular velocity on $\ec$ and $\kappa$ is less at high temperatures than that at low temperatures (figures \ref{Fig.1}a and \ref{Fig.1}b), which can be understood from the fact that, at high temperatures, the energy scale associated with the temperature dominates over the energy scale related to the angular velocity, thus suppressing the impact of angular velocity on conductivities. In the novel relaxation time approximation, a decrease in $\ec$ and $\kappa$ is observed as compared to the frequently used relaxation time approximation (figures \ref{Fig.1}a and \ref{Fig.1}b). The difference between the results in these two approximations is mainly attributed to the different relaxation times and different collision integrals in both cases. It is also found that the distinction between the RTA and the novel RTA is more conspicuous at higher temperatures both 
for the nonrotating and rotating QGP mediums. On the whole, the rotation facilitates the transport of charge and heat in the medium. 

We have also compared our results with the results obtained from other models, such as the lattice QCD \cite{Gupta:PLB597'2004,Ding:LATTICE2014'2015,Aarts:JHEP02'2015}, the Nambu-Jona-Lasinio (NJL) model \cite{Marty:PRC88'2013}, the effective fugacity quasiparticle (EFQP) model \cite{Mitra:PRD96'2017} and the extended linear sigma (ELS) model \cite{Tawfik:AN20240004'2024}. We note that the aforesaid models including the lattice simulation with the rotation are at an early stage of development, and the results on electrical and thermal conductivities at finite rotation are not yet available in these models, so we have used the results in the absence of rotation from these models. The electrical conductivity in the hot phase of the QCD plasma is extracted from a quenched lattice measurement of the Euclidean time vector correlator for $1.5\leq T/T_c \leq 3$ in ref. \cite{Gupta:PLB597'2004}. This study estimates the value of the electrical conductivity that lies above our results at zero angular velocity as well as at finite angular velocity. In ref. \cite{Ding:LATTICE2014'2015}, the electrical conductivity has been calculated using the vector correlation function within the quenched lattice QCD and this result is found to be in close proximity to our results throughout the considered temperature range. Lattice result in ref. \cite{Aarts:JHEP02'2015} lies slightly below our result calculated using the RTA method, whereas it is closer to our result determined using the novel RTA method. Upon comparing our results with the NJL model \cite{Marty:PRC88'2013}, EFQP model \cite{Mitra:PRD96'2017} and ELS model \cite{Tawfik:AN20240004'2024} calculations, it is found that the results of the aforesaid models lie above our results on electrical conductivity in the absence as well as in the presence of finite angular velocity. However, as compared to the EFQP model estimation, the NJL model and the ELS model estimations on electrical conductivity are in close agreement with our results at low temperatures. We note that the lattice QCD calculations on the thermal conductivity are not available in literature, so we have only compared our result on the thermal conductivity with that in the NJL model \cite{Marty:PRC88'2013}, EFQP model \cite{Mitra:PRD96'2017} and ELS model \cite{Tawfik:AN20240004'2024}. We have found that the results from these models lie above our result over the entire range of temperature for both nonrotating and rotating mediums. However, at high temperatures, our result on the thermal conductivity is closer to the estimations 
of these models. 

\begin{figure}[]
\begin{center}
\begin{tabular}{c c}
\includegraphics[width=7.4cm]{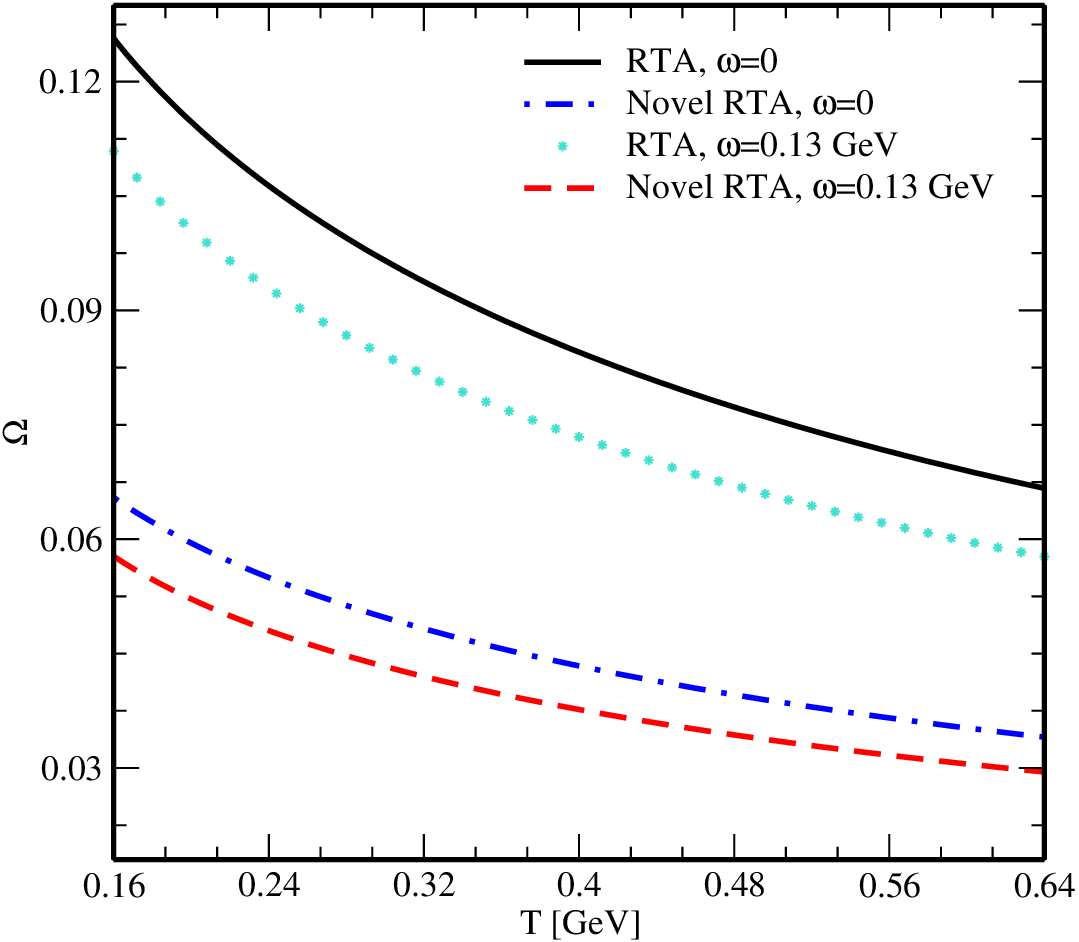}&
\hspace{0.73 cm}
\includegraphics[width=7.4cm]{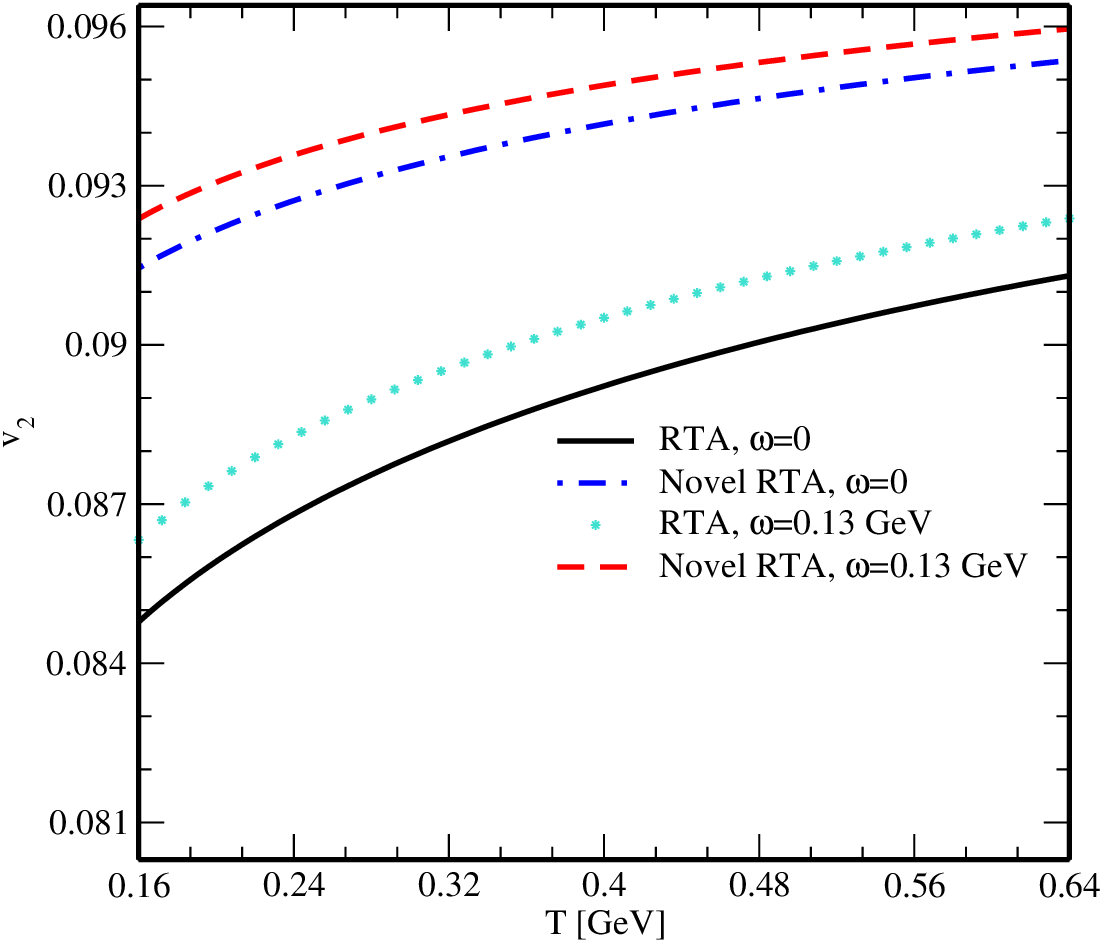} \\
a & b
\end{tabular}
\caption{Variations of (a) the Knudsen number and (b) the elliptic flow with temperature for different values of angular velocity.}\label{Fig.2}
\end{center}
\end{figure}

In figure \ref{Fig.2}, the effects of rotation on the Knudsen number ($\Omega$) and the elliptic flow ($v_2$) have been displayed. Observation shows a decrease in the Knudsen number (figure \ref{Fig.2}a) and an increase in the elliptic flow (figure \ref{Fig.2}b) when rotation is introduced in the medium. This behavior of the Knudsen number at finite angular velocity is corroborated by the observations on the thermal conductivity and the specific heat at constant volume in the similar regime. The increase of $v_2$ at finite rotation also agrees with the observation in ref. \cite{Becattini:PRC77'2008}. Thus the medium approaches towards the equilibrium with the increase of rotation. Further, the novel relaxation time approximation estimates smaller value of $\Omega$ and larger value of $v_2$ as compared to their corresponding values in the relaxation time approximation. This indicates an enhancement of the interactions among the produced particles in heavy ion collisions and faster thermalization of the medium. 

\begin{figure}[]
\begin{center}
\begin{tabular}{c c}
\includegraphics[width=7.4cm]{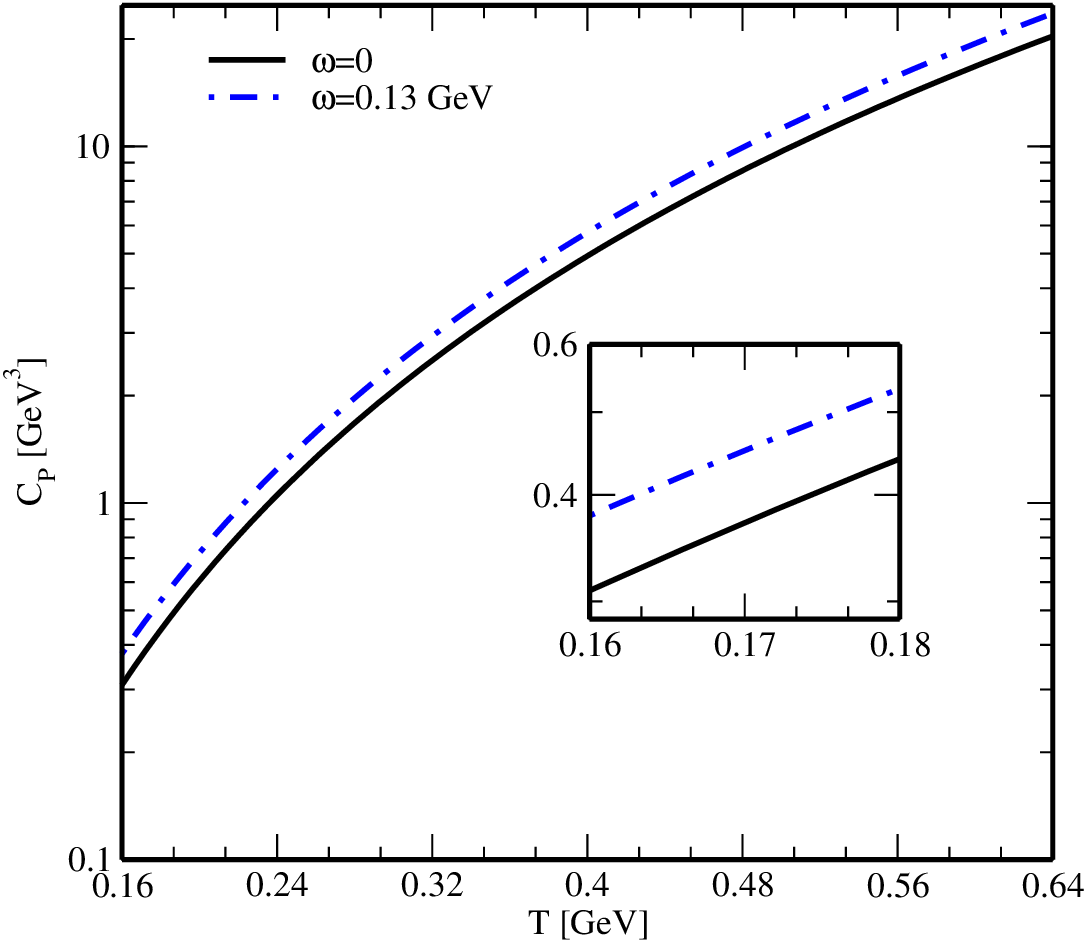}&
\hspace{0.73 cm}
\includegraphics[width=7.4cm]{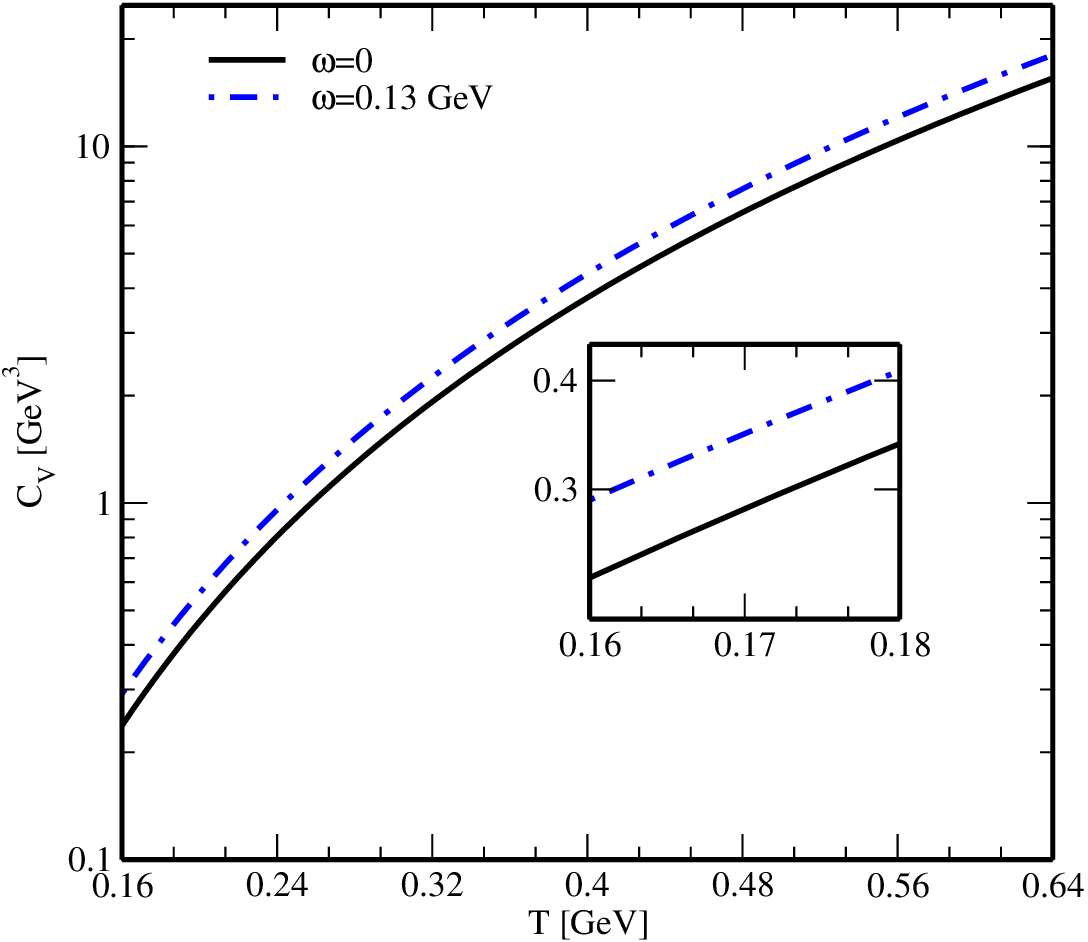} \\
a & b
\end{tabular}
\caption{Variations of (a) the specific heat at constant pressure and (b) the specific heat at constant volume 
with temperature for different values of angular velocity.}\label{Fig.3}
\end{center}
\end{figure}

\begin{figure}[]
\begin{center}
\begin{tabular}{c c}
\includegraphics[width=7.4cm]{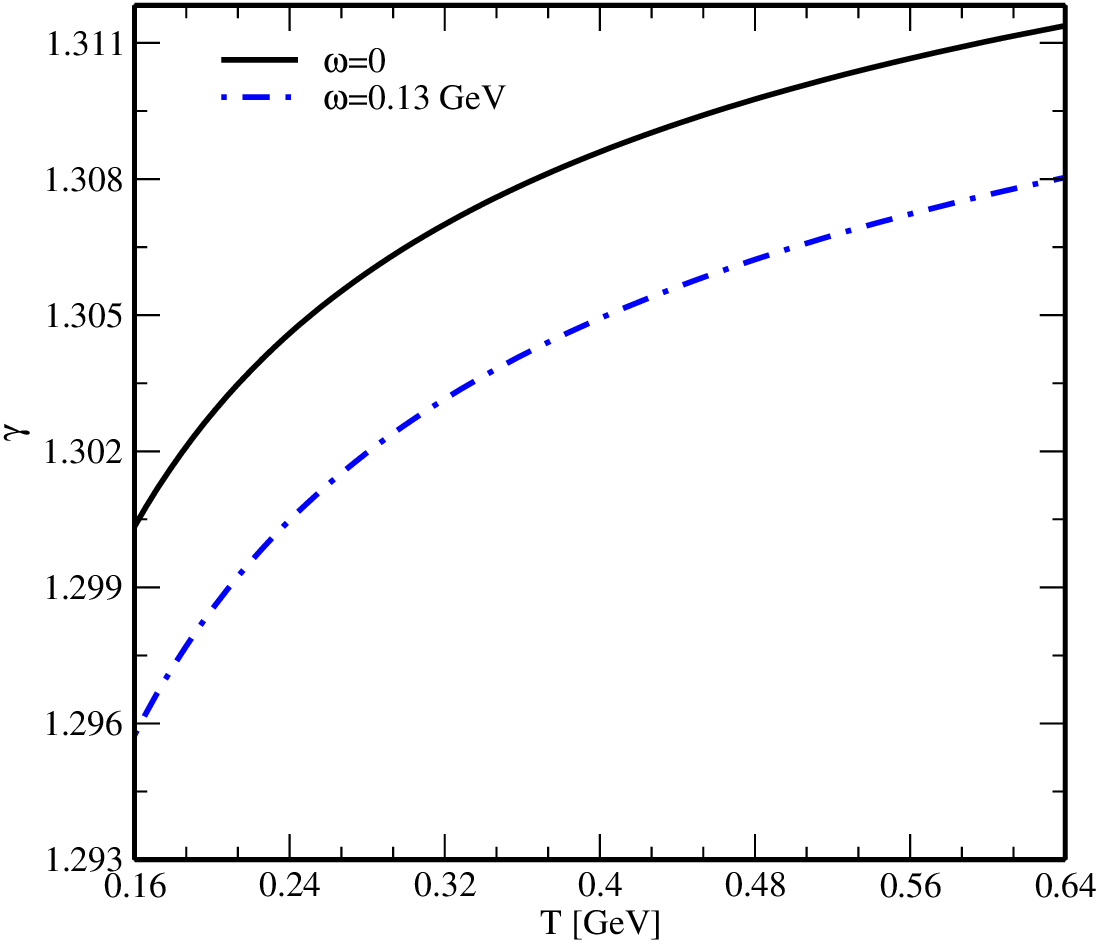}&
\hspace{0.73 cm}
\includegraphics[width=7.4cm]{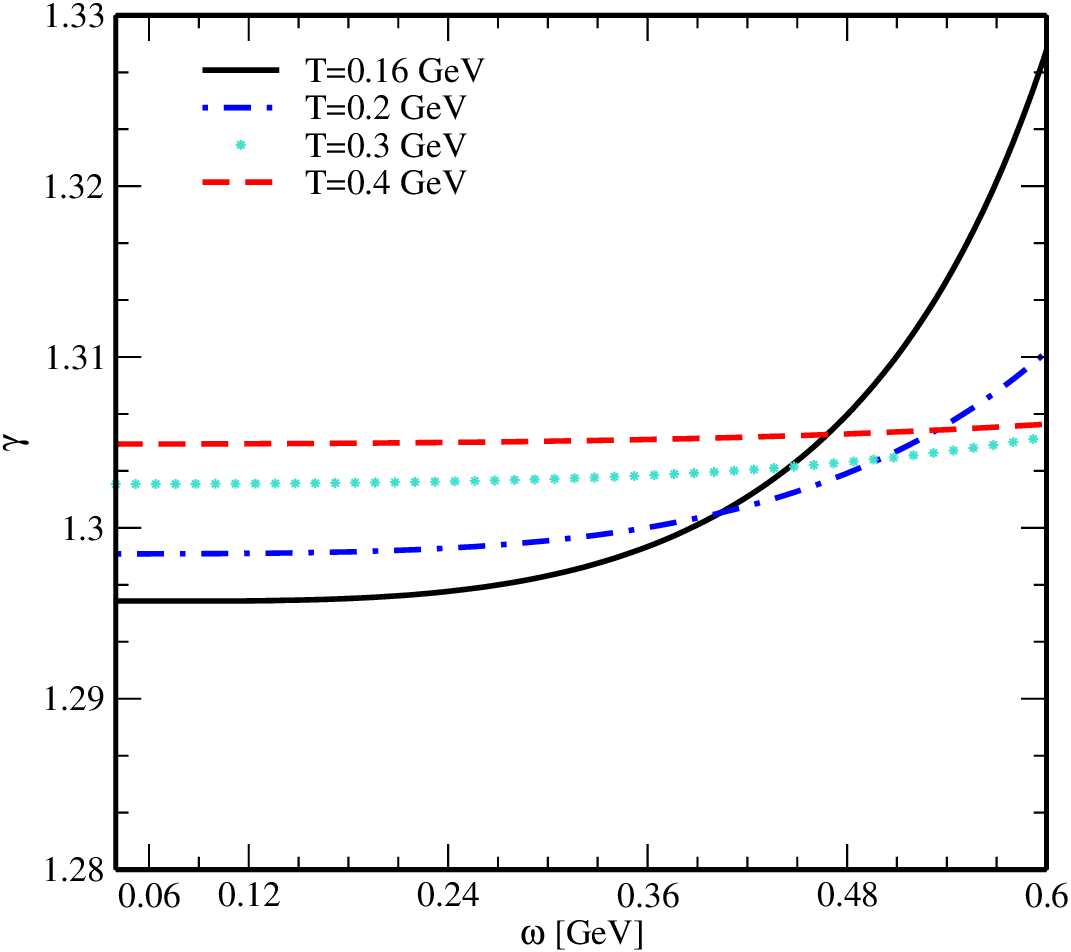} \\
a & b
\end{tabular}
\caption{Variations of the adiabatic index ($\gamma=C_P/C_V$) (a) with temperature for different values of angular velocity and (b) with angular velocity for different values of temperature.}\label{Fig.4}
\end{center}
\end{figure}

Figure \ref{Fig.3} shows the effects of rotation on the specific heat at constant pressure ($C_P$) and the specific heat at constant volume ($C_V$) over a given range of the temperature. It is observed that the emergence of rotation enhances both $C_P$ and $C_V$, with the magnitude of the former being larger than that of the latter for all temperatures. Thus, for a rotating medium, the changes in enthalpy and energy density with temperature are larger as compared to the nonrotating case. The relative behavior between $C_P$ and $C_V$ can be discerned through their ratio $\gamma$ or the adiabatic index. Figure \ref{Fig.4} displays how the adiabatic index varies with the temperature and with the angular velocity. It is observed that the adiabatic index becomes larger as the medium gets hotter, whereas, if the rotation is introduced into the medium, $\gamma$ gets decreased as compared to the $\omega=0$ case (figure \ref{Fig.4}a). On the other hand, within the rotating medium, $\gamma$ is found to slowly increase with the increase of angular velocity. Further, the trend of variation of $\gamma$ with the angular velocity is not same for all temperatures, rather the increase of $\gamma$ with angular velocity becomes faster for lower temperatures (figure \ref{Fig.4}b). In low angular velocity regime, the variation of $\gamma$ is meagre, however, a noticeable increasing behavior of $\gamma$ is observed in high angular velocity regime. This indicates that the change in enthalpy is larger than the change in energy density for an increase of one unit in temperature and this difference gets enhanced as the rotation of the QGP medium becomes more rapid. 

\begin{figure}[]
\begin{center}
\begin{tabular}{c c}
\includegraphics[width=7.4cm]{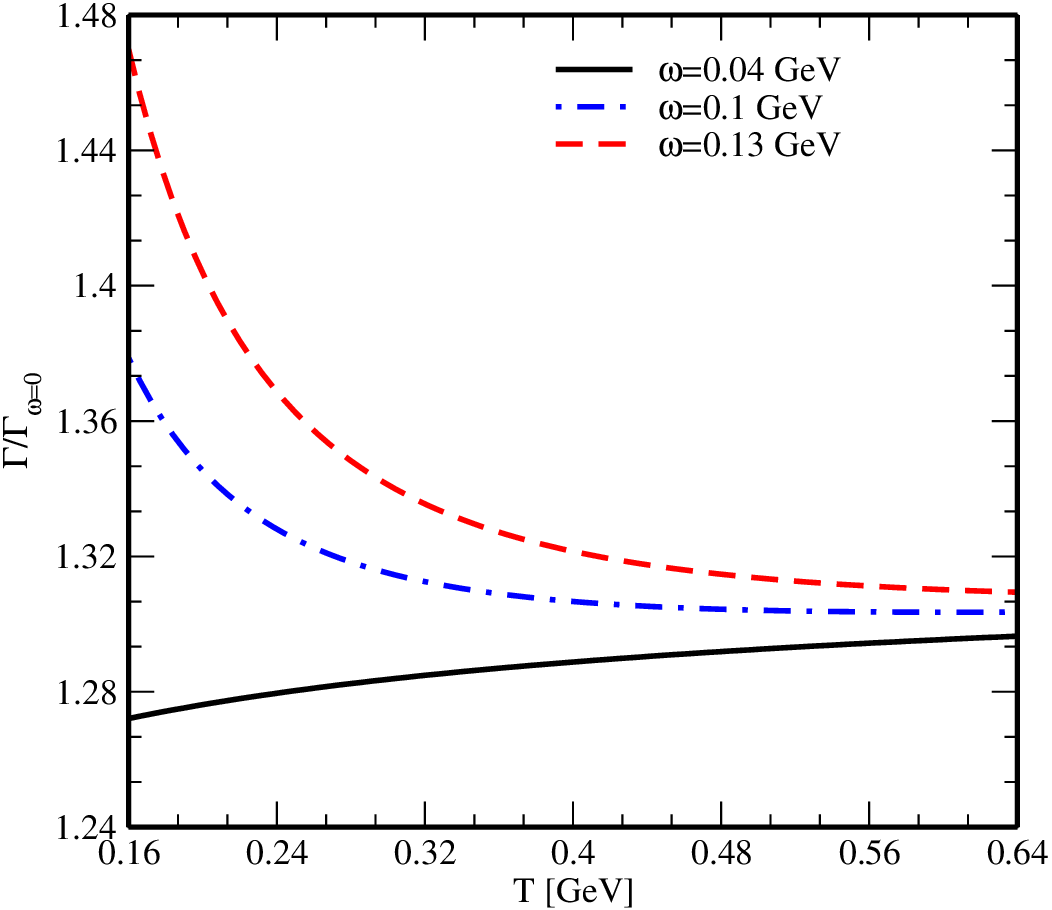}&
\hspace{0.73 cm}
\includegraphics[width=7.4cm]{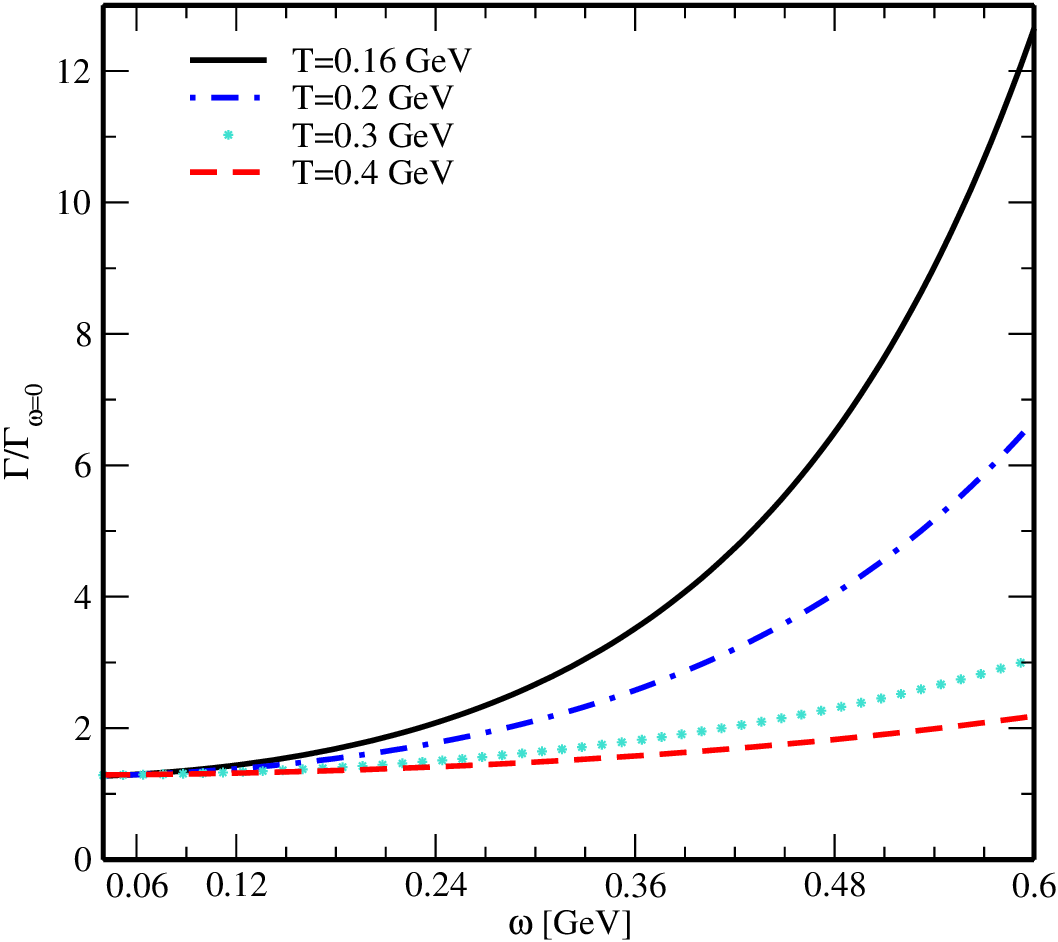} \\
a & b
\end{tabular}
\caption{Variations of the trace anomaly scaled with its value in the absence of rotation (a) as a function 
of temperature for different values of angular velocity and (b) as a function of angular velocity for 
different values of temperature.}\label{Fig.5}
\end{center}
\end{figure}

In figure \ref{Fig.5}, the variations of the trace anomaly in units of its value in the absence of rotation with temperature and with angular velocity have been depicted. It is observed that, with the increase 
of temperature, the deviation of the trace anomaly at finite angular velocity from that at zero angular velocity decreases (figure \ref{Fig.5}a). Thus, at high temperatures, rotation of the medium has a meagre impact on the trace anomaly. Our observation has further found an increasing trend of the trace anomaly as the angular velocity increases and the slope decreases if the medium becomes hotter (figure \ref{Fig.5}b). The effect of angular velocity on the trace anomaly seems to be more pronounced at lower temperatures. Thus the rotation makes the constituents of the medium more interactive and as a result, it takes the medium a bit away from the conformal symmetry. 

\begin{figure}[]
\begin{center}
\begin{tabular}{c c}
\includegraphics[width=7.4cm]{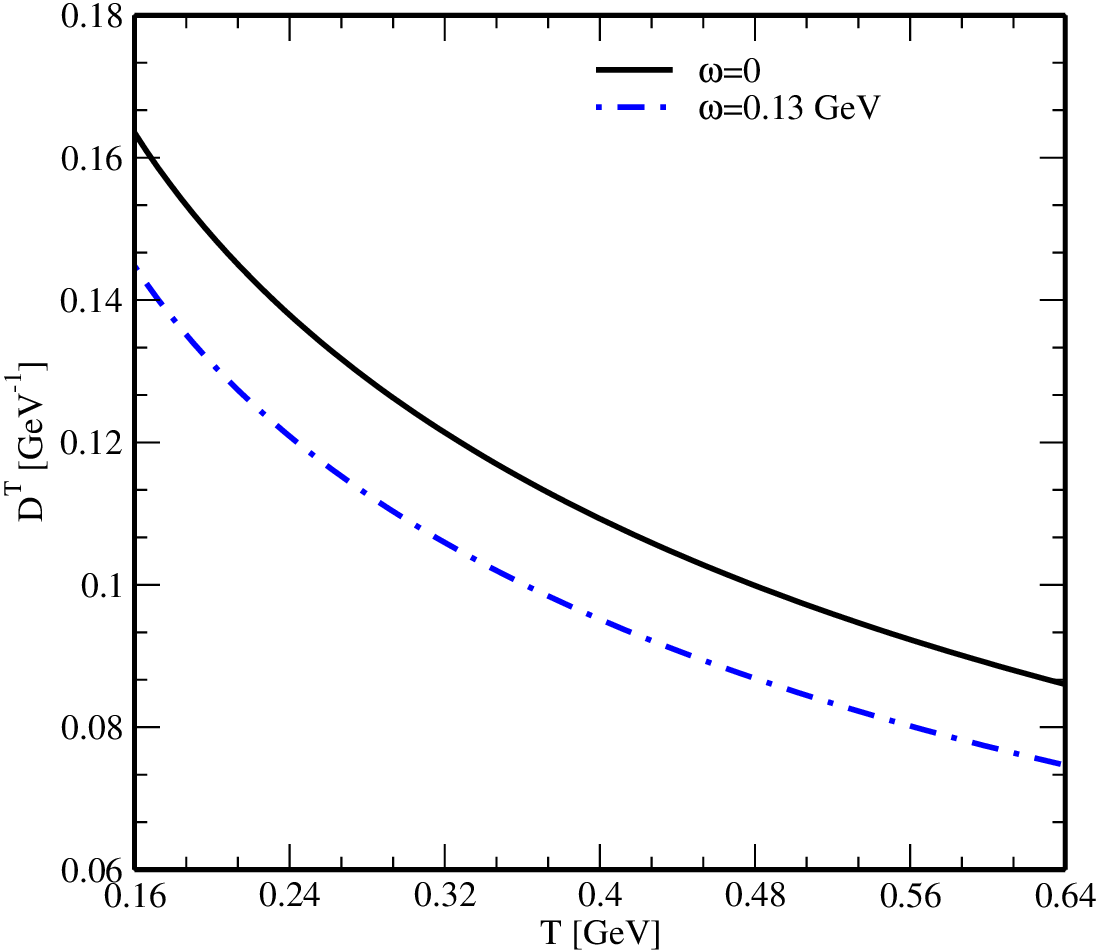}&
\hspace{0.73 cm}
\includegraphics[width=7.4cm]{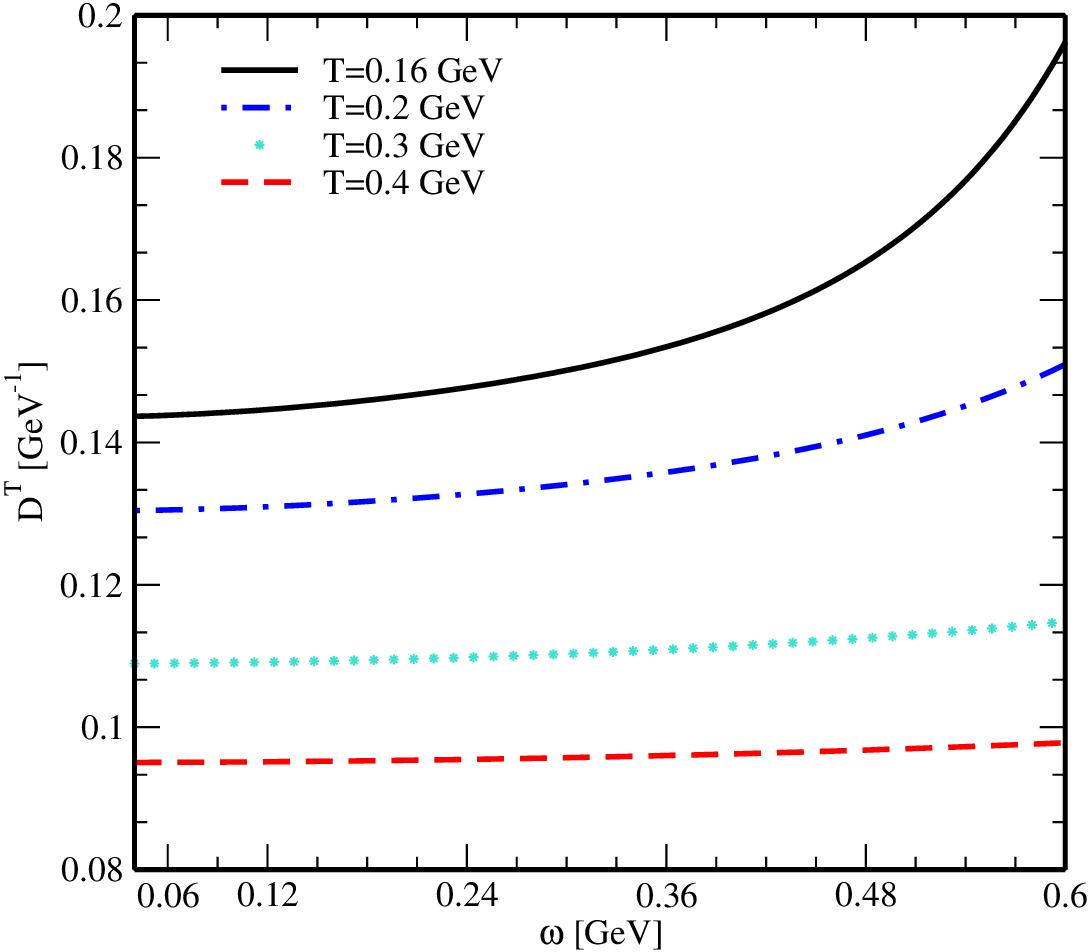} \\
a & b
\end{tabular}
\caption{Variations of the thermal diffusion constant (a) as a function of temperature for different values of angular velocity and (b) as a function of angular velocity for different values of temperature.}\label{Fig.6}
\end{center}
\end{figure}

\begin{figure}[]
\begin{center}
\begin{tabular}{c c}
\includegraphics[width=7.4cm]{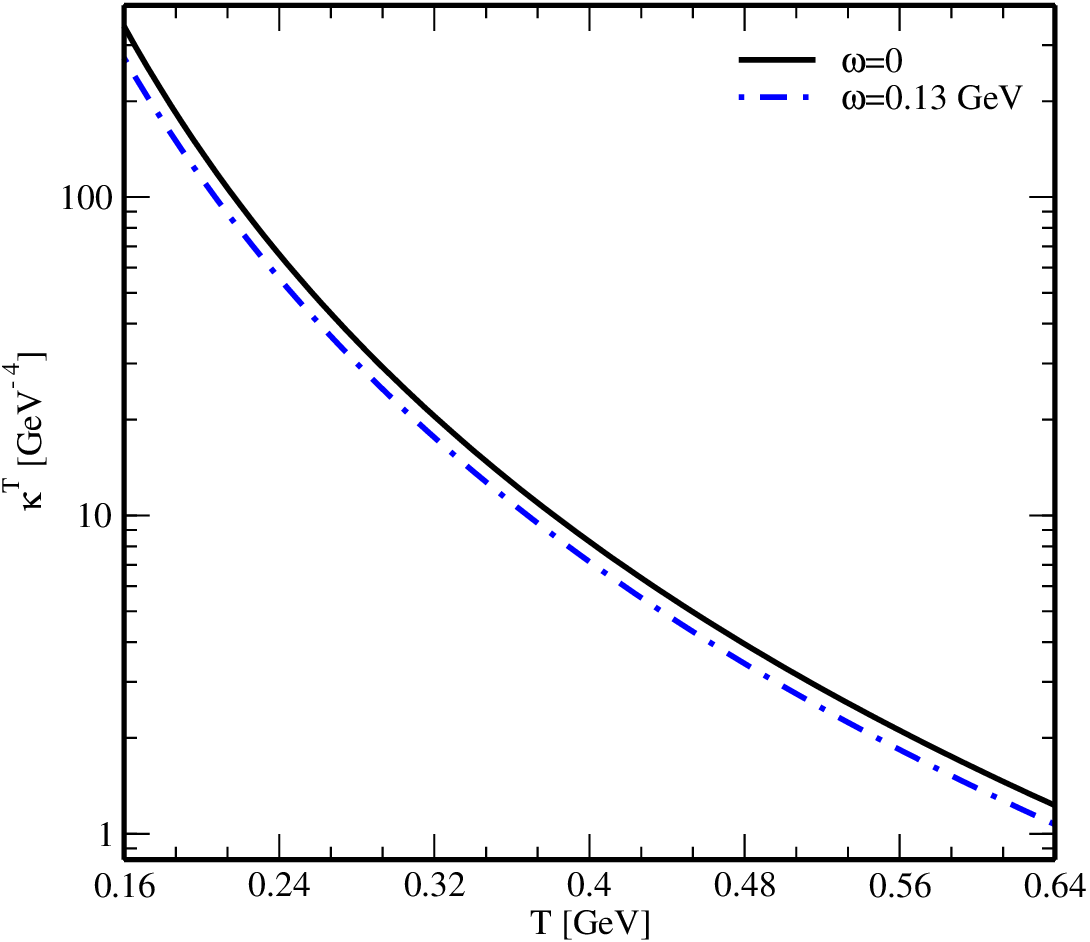}&
\hspace{0.73 cm}
\includegraphics[width=7.4cm]{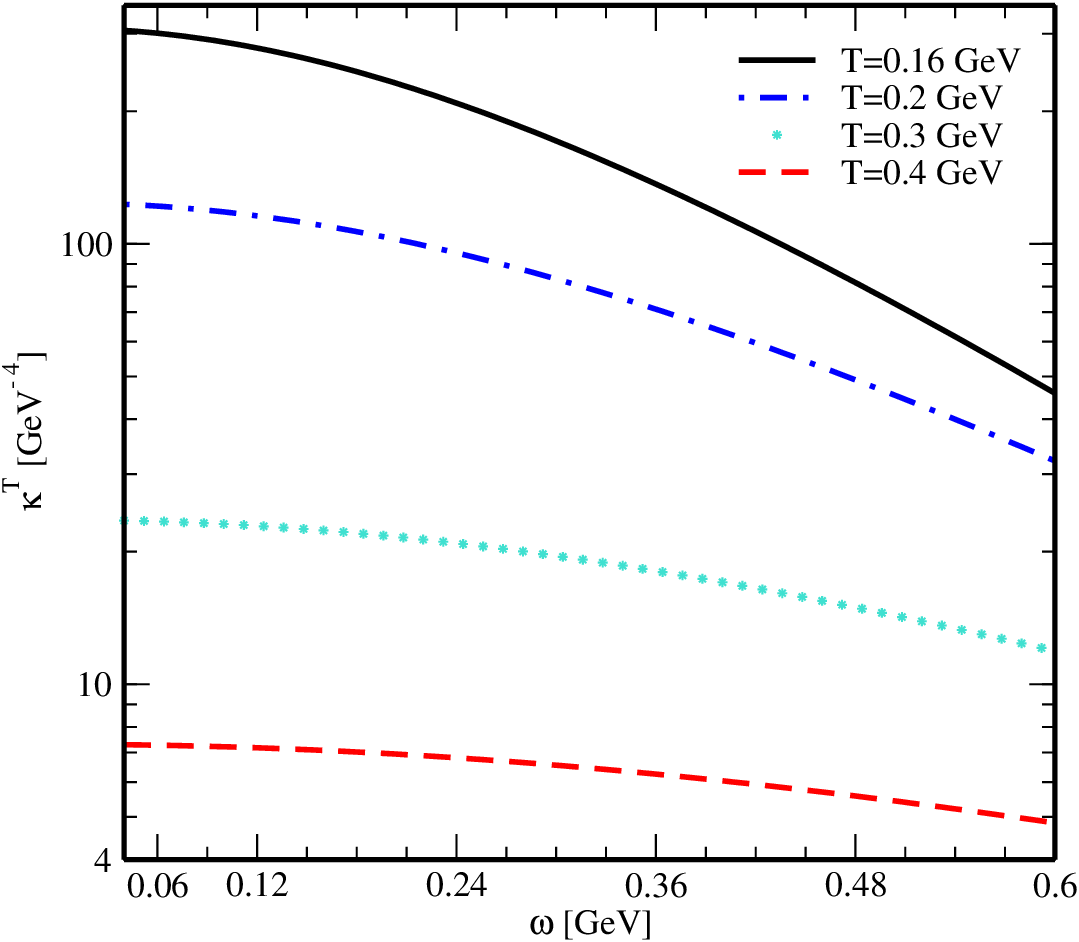} \\
a & b
\end{tabular}
\caption{Variations of the isothermal compressibility (a) as a function of temperature for different values of angular velocity and (b) as a function of angular velocity for different values of temperature.}\label{Fig.7}
\end{center}
\end{figure}

Figures \ref{Fig.6} and \ref{Fig.7} respectively display how the thermal diffusion constant ($D^T$) and the isothermal compressibility ($\kappa^T$) get modulated by the rotation with varying temperature. It is 
observed that the rotation of the medium reduces the thermal diffusion constant over the entire range of temperature (figure \ref{Fig.6}a). This reduction in $D^T$ is mainly attributed to the variations of $\kappa$ and $C_P$ in the similar regime. With the increase of angular velocity, the increase in the specific heat at constant pressure dominates over the increase in the thermal conductivity, therefore, a decrease in the thermal diffusion constant at finite angular velocity is observed. Smaller value of the thermal diffusion constant indicates slower heat transfer in a rotating medium as compared to that in a nonrotating medium. On the other hand, within the rotating medium, $D^T$ is found to slowly increase with the increase of angular velocity, which is evident only in the regime of low temperature and high angular velocity (figure \ref{Fig.6}b), where the energy scale associated with the angular velocity prevails over the energy scale related to the temperature. As the temperature increases, the isothermal compressibility is found to decrease (figure \ref{Fig.7}a). Further, the introduction of angular velocity also reduces $\kappa^T$ and the matter may show nearly perfect fluid behavior. The rate of decline is larger in high angular velocity regime than in low angular velocity regime, whereas the rate of decline gets gradually smaller as the rotating medium gets hotter (figure \ref{Fig.7}b). This suggests that, with the increasing angular velocity, it is difficult to 
compress the system. Thus, it is inferred that the matter becomes stiffer as the rotation becomes more rapid. 

\section{Conclusions}
In this work, we explored the transport coefficients and observables of a rotating QGP medium. We calculated the electrical and the thermal conductivities using the relativistic Boltzmann transport equation 
in the novel relaxation time approximation within the kinetic theory approach. We also compared our results with the frequently used relaxation time approximation. We observed that the introduction of rotation enhances the conduction of charge and heat in the medium. However, the rotation leaves a modest impact on the conductivities. In addition, the novel relaxation time approximation estimates smaller values of the electrical and the thermal conductivities as compared to their corresponding values 
in the relaxation time approximation. We further explored the effects of rotation on various observables, such as the Knudsen number, the elliptic flow, the specific heat at constant pressure, the specific heat at constant volume, the trace anomaly, the thermal diffusion constant and the isothermal compressibility. From the behaviors of these observables, we found that the rotation leaves a moderate influence on the equilibrium property of the medium, the interactions among particles produced in heavy ion collisions, the adiabatic 
index, the conformal symmetry, the rate of the heat transfer in the medium and the stiffness of the medium. 

\section*{Acknowledgments}
One of the present authors (S. R.) acknowledges financial support from ANID Fondecyt Postdoctoral Grant 3240349 and S. D. acknowledges the SERB Power Fellowship, SPF/2022/000014 for the support on this work.

\end{document}